# Chiroptical Ternary Entropy Harvesting from Self-Assembled Block Copolymer Nanopatterns

*Wookjin Jung[1,#], Serin Jeong[1,#], Kyulim Kim[1,#], Dongkyu Lee[1], Sang Ouk Kim[1,4,*], and Jihyeon Yeom[1,2,3,4,*].*

[1] Department of Materials Science and Engineering, Korea Advanced Institute of Science and Technology (KAIST), 291 Daehak-ro, Yuseong-gu, Daejeon 34141, Republic of Korea

[2] Department of Biological Sciences, Korea Advanced Institute of Science and Technology (KAIST), 291 Daehak-ro, Yuseong-gu, Daejeon 34141, Republic of Korea

[3] Institute for Health Science and Technology, Korea Advanced Institute of Science and Technology (KAIST), 291 Daehak-ro, Yuseong-gu, Daejeon 34141, Republic of Korea

[4] Institute for the NanoCentury, Korea Advanced Institute of Science and Technology (KAIST), 291 Daehak-ro, Yuseong-gu, Daejeon 34141, Republic of Korea

[*] Corresponding author

[#] Authors contributed equally

**ABSTRACT:**

Nanoscale fabrication inevitably produces local stochasticity that is commonly treated as a defect, but can instead be harnessed as a material resource for information security. Here we report a chiroptical platform for ternary entropy harvesting based on stochastic Au nanopatterns formed by block copolymer self-assembly. By transducing fabrication-induced stochastic microstates into handedness-dependent optical responses through Raman optical activity mapping, our platform enables native ternary digitization rather than conventional binary encoding, allowing physically harvested ternary random sequences to be used for key generation. This raises the information density to $\log_2 3 = 1.585$ bits per trit, approximately 58.5% higher than the binary limit, enabling more entropy to be harvested from a limited physical footprint. The harvested outputs exhibit near-balanced symbol populations, negligible spatial and inter-sample correlations, Shannon entropy approaching the ternary ideal, and resistance to statistical and machine-learning-based prediction. These results establish self-assembled chiroptical nanostructures as a scalable platform for cryptographic key generation, secure edge devices, and distributed Internet-of-Things platforms.

**KEYWORDS:** chiral nanomaterials, block copolymer self-assembly, defect-driven nanostructures, nanoplasmonics

Randomness is the stubborn residue of nanomaterials fabrication. Even when the governing interactions are well understood, unavoidable local stochasticity—including thermal fluctuations, kinetic trapping and microscopic disorder—can drive a system into one of many accessible microstates.[1-3] The result is run-to-run morphological variability that is typically classified as a defect. Accordingly, much of modern nanofabrication has focused on suppressing stochastic deviations in pursuit of deterministic structure control, a natural objective in precision manufacturing.[3-5] From the standpoint of statistical physics and information science, however, irreducible variability is not merely a failure of control but a potential source of beneficial functionalities.[6-8] If such stochastic microstates can be transduced into measurable digital outputs, the same randomness that limits deterministic nanomanufacturing can instead be repurposed as a materials resource for information security, based on random number generation.

Among nanoscale fabrication strategies, block copolymer (BCP) self-assembly is particularly attractive in this regard because it naturally generates dense nanopatterns with rich local morphological variability while remaining compatible with large-area, scalable processing.[9-11] In BCP systems, microphase separation and self-organization produce nanoscale patterns whose local orientations and geometries are shaped by unavoidable stochasticity during assembly. Although such variability is usually treated as an imperfection from the standpoint of pattern regularity, it also provides a physically embedded source of structural complexity. Crucially, self-assembly is not merely a low-cost route to nanopatterning here; it defines the physical origin of the entropy source itself. Each independently self-assembled substrate realizes a distinct stochastic morphological landscape, collectively furnishing a practically vast library of entropy-bearing nanostructures.

Harnessing this materials-intrinsic randomness requires a readout modality that is simultaneously sensitive to local nanoscale heterogeneity, compatible with large-area sampling and capable of converting structural stochasticity into statistically robust digital symbols. Most physical random number generators rely on electrical noise, quantum fluctuations or device-level stochastic switching.[12-15] By contrast, materials-driven entropy harvesting remains less explored, particularly in self-assembled nanostructures, where stochastic patterned structures are abundant but difficult to transduce directly into information-rich outputs. A particularly appealing opportunity arises when the readout observable is inherently multi-level rather than binary, because this allows more of the structural information encoded in the material to be retained during digitization.

Chiroptical interactions provide such a route. Even when a nanostructured surface is globally achiral on average, subtle local geometric asymmetries can generate handedness-dependent electromagnetic responses under circularly polarized illumination.[16-21] Because both the sign and magnitude of such responses depend sensitively on the local nanoscale arrangement,[22,23] chiroptical observables offer a natural way to probe stochastic structural heterogeneity. In particular, a signed chiroptical response can be digitized natively into more than two states, opening a pathway to multi-level entropy harvesting without artificial encoding.[24,25] Despite substantial advances in chiral plasmonics and metasurfaces, chiroptical readout has rarely been exploited as a materials transduction strategy for harvesting entropy from self-assembled stochastic nanostructures.

Here we introduce a chiroptical platform for ternary entropy harvesting based on stochastic Au nanopatterns replicated from block copolymer self-assembly. The self-assembled morphology is transferred into plasmonic Au patterns whose local electromagnetic response depends on the handedness of circularly polarized excitation. We probe this stochastic chiroptical landscape using Raman optical activity (ROA) mapping, which measures the differential Raman response under right- and left-circularly polarized illumination with sub-micron optical localization. Because the resulting chiroptical contrast is intrinsically signed, it can be digitized naturally into three classes—negative, neutral and positive—yielding native ternary symbols rather than a forced binary output. In this framework, the self-assembled nanomaterial is not simply a passive substrate for a separate randomness source, but the entropy-bearing material itself.

A key advantage of this approach is that it combines locally rich structural disorder with statistically well-behaved ensemble-level output. Across large-area ROA maps, the measured local responses form a stable distribution centered near zero, with positive and negative tails corresponding to opposite handedness-dominant optical environments. By applying symmetric thresholds to this distribution, we convert the spatially resolved chiroptical response into trits, increasing the symbol-level information capacity to $\log_2 3 = 1.585$ bits per trit, approximately 58.5% higher than binary encoding for the same number of sampled sites. Importantly, the harvested entropy does not rely on external circuitry or algorithmic post-processing but emerges directly from fabrication-induced stochastic microstates encoded in the nanomaterial. These results show that structural disorder in nanomaterials, conventionally regarded as an unwanted defect to be minimized, can instead be harnessed as a functional resource for information security.

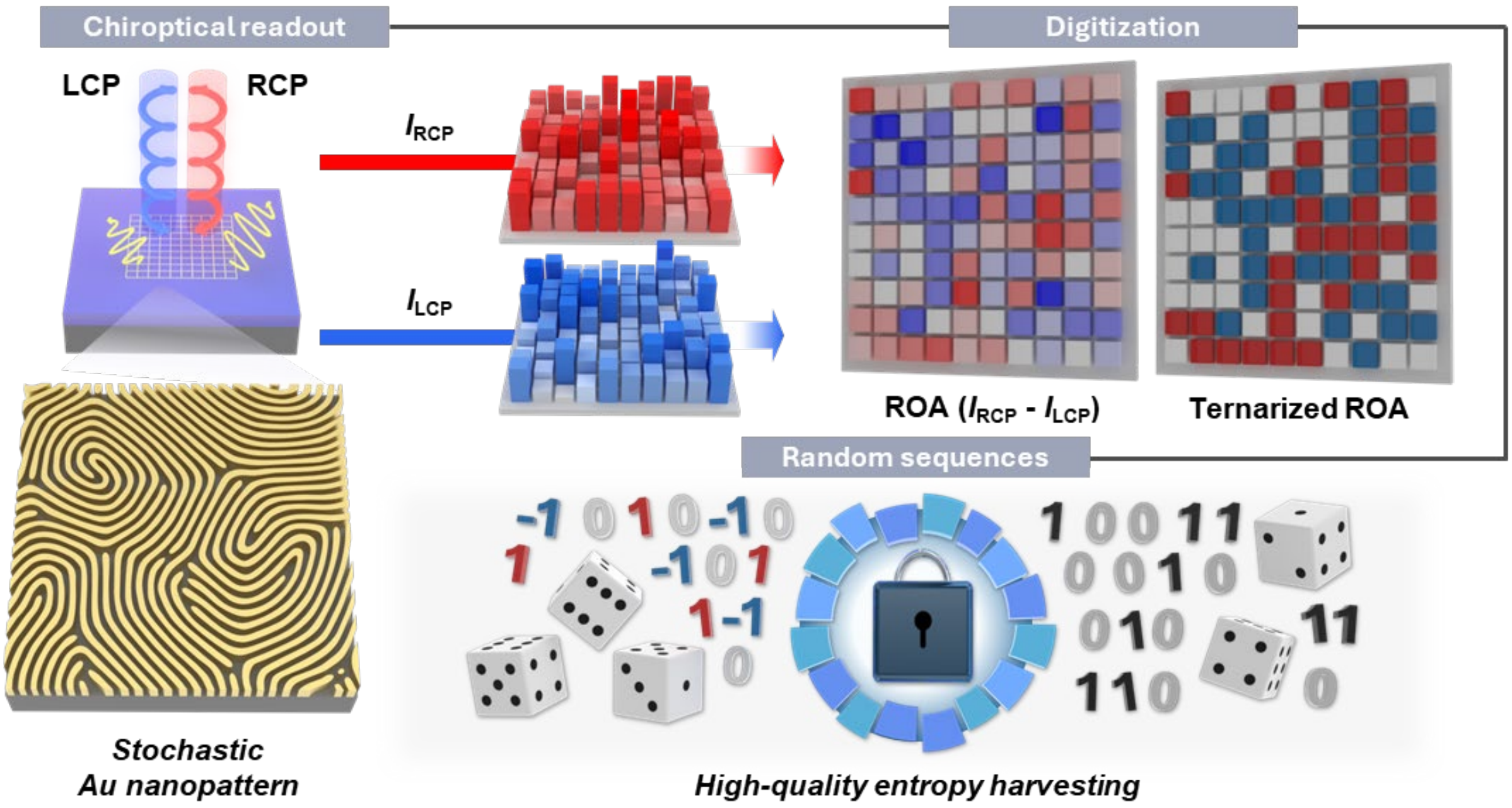

**Scheme 1.** Entropy harvesting through ternary random sequence generation using Raman optical activity mapping of stochastic Au nanopatterns formed by block copolymer self-assembly.

## Results and Discussion

To obtain nanopatterns for the ternary entropy harvesting, we spin-coated a solution of polystyrene-*block*-poly(methyl methacrylate) (PS-*b*-PMMA) onto a Si wafer, followed by solvent and/or thermal annealing to induce microphase separation. After self-assembled pattern formation, the PMMA domains were selectively removed by etching, creating nanoscale topographic patterns that replicate original BCP morphology. Au was deposited over the entire patterned area to transform into Au nanopatterns by lift-off process. A thin Cr layer (~2 nm) was deposited prior to Au deposition to serve as an adhesion layer between the Au and the underlying Si substrate **(Figure 1A)**.[26]

Scanning electron microscopy (SEM) confirmed that the fabricated nanopatterns exhibit a stochastic, fingerprint-like morphology **(Figure 1B)**. Specifically, intrinsic immiscibility between the two blocks drives microphase separation into lamellar nanostructures.[27,28] Without external pattern guidance, the lamellar nanodomains adopt random in-plane orientations with localized curvatures, producing the characteristic fingerprint-like appearance.[29]

We next examined the out-of-plane structure of Au nanopatterns by cross-sectional transmission electron microscopy (TEM) **(Figure 1C)**. Well-defined, vertical wall-like features were observed with an average lateral period of ~64 nm. The thickness of the patterned metal layer was ~11.6 nm. To further resolve the compositional stratification, we performed energy-dispersive X-ray spectroscopy (EDS) line profiling across the film stack **(Figure 1D)**. The EDS results confirm the formation of a ~2.4 nm Cr adhesion layer on the Si substrate, overlaid by a ~9.2 nm Au layer, consistent with the TEM analysis.

To statistically characterize the structural ordering of the Au nanopatterns over a large area, grazing-incidence small-angle X-ray scattering (GISAXS) measurements were performed **(Figure 1E, F)**. The well-resolved Bragg rods in the 2D scattering pattern demonstrated excellent lateral ordering of the Au lines over a macroscopic area. The corresponding 1D in-plane cut reveals a series of peaks with relative positions of 1:2:3:4, confirming the formation of a periodic line grating structure, a typical characteristic of the lamellar morphology. Based on the position of the first-order peak (q*), the nanopattern periodicity (center-to-center spacing) was determined to be 64.78 nm, in good agreement with the value obtained from cross-sectional TEM.

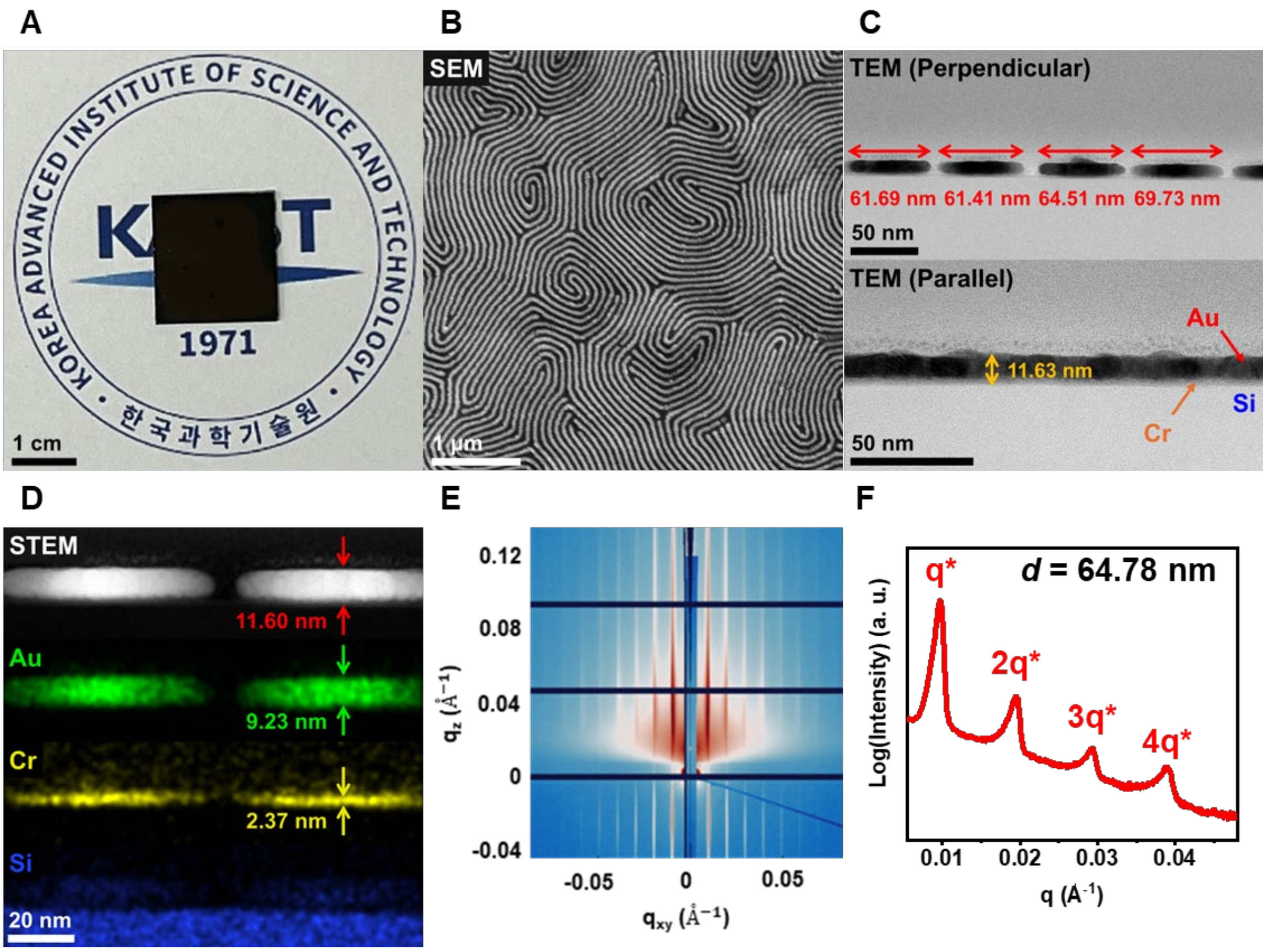

**Figure 1.** Structural characterization of Au nanopatterns. (A) Photograph of the fabricated sample. (B) Top-down SEM image showing fingerprint-like morphology. (C) Cross-sectional TEM images and (D) corresponding EDS elemental mapping results showing the distribution of Au, Cr, and Si. (E) 2D GISAXS scattering pattern and (F) the corresponding 1D in-plane intensity profile.

We could further control the nanopattern dimensions by changing the molecular weight of the PS-*b*-PMMA (**Figure S1**). The center-to-center spacing (pitch) between neighboring lamellae was readily adjusted depending on the polymer chain length. Specifically, a low molecular weight (49.5k-*b*-48.5k) yielded a reduced pitch of 43.01 nm, while a higher molecular weight (170k-*b*-178k) increased the pitch to 89.63 nm. Regardless of the periodicity, all the lamellar nanopatterns preserved their intrinsic stochastic, fingerprint-like morphology.[30,31] We exploited this inherent randomness, specifically the local asymmetry of the fabricated Au nanopatterns, as the physical entropy source for the entropy harvesting. To read out this local chirality, we employed ROA, which measures the differential Raman scattering intensity produced by right-circularly polarized incident light (RCP) and left-circularly polarized incident light (LCP) and serves as a probe of chiral asymmetry. Here, ROA denotes a Raman circular intensity-difference signal generated by handedness-dependent plasmonic near-field enhancement.

ROA is defined as the difference between the Raman emission intensities measured under RCP and LCP excitation. Under the electromagnetic enhancement approximation for plasmonic Raman scattering (**Supplementary Notes**),[32] the ROA signal at position $\mathbf{r}$, $I_{\mathrm{ROA}}(\mathbf{r})$ can be expressed as

$$I_{\mathrm{ROA}}(\mathbf{r}) = I_{\mathrm{RCP}}(\mathbf{r}) - I_{\mathrm{LCP}}(\mathbf{r}) \propto |E_{\mathrm{RCP}}(\mathbf{r})|^4 - |E_{\mathrm{LCP}}(\mathbf{r})|^4 \equiv \Delta|E(\mathbf{r})|^4 \quad (1)$$

, where $I_{\mathrm{RCP}}(\mathbf{r})$ and $I_{\mathrm{LCP}}(\mathbf{r})$ denote the Raman emission intensities measured under RCP and LCP excitation, respectively, and $|E_{\mathrm{RCP}}(\mathbf{r})|$ and $|E_{\mathrm{LCP}}(\mathbf{r})|$ denote the corresponding local electric near-field amplitudes generated under RCP and LCP illumination at position $\mathbf{r}$. Unlike circular dichroism spectroscopy, which typically averages chirality over a macroscopic (centimeter-scale) area, ROA utilizes a focused laser beam with a diameter of several hundred nanometers.[33,34] This feature makes it an ideal tool for resolving local chirality at the nanoscale.[35]

Prior to experimental validation, we theoretically assessed the handedness-dependent near-field asymmetry of the Au nanopatterns using finite-element simulations. Realistic 3D models of the Au nanopatterns were reconstructed by binarizing experimentally obtained high-resolution SEM images of the samples (**Figure S2**). Using the 3D models, the nanopattern-induced local electric near fields were simulated by solving Maxwell's equations in COMSOL Multiphysics under 633 nm circularly polarized illumination. Right- and left-circularly polarized excitations were applied using identical incident amplitude and illumination geometry to isolate handedness-dependent local electric near-field asymmetry. At each spatial coordinate $\mathbf{r}$, we extracted the electric near-field magnitude under RCP and LCP excitation, $|E_{\mathrm{RCP}}(\mathbf{r})|$ and $|E_{\mathrm{LCP}}(\mathbf{r})|$, and defined the differential near-field map as $\Delta E(\mathbf{r}) = |E_{\mathrm{RCP}}(\mathbf{r})| - |E_{\mathrm{LCP}}(\mathbf{r})|$ (**Figure S3A, B**). We further mapped a normalized electric near-field dissymmetry factor, $g(\mathbf{r}) = \frac{|E_{\mathrm{RCP}}(\mathbf{r})| - |E_{\mathrm{LCP}}(\mathbf{r})|}{|E_{\mathrm{RCP}}(\mathbf{r})| + |E_{\mathrm{LCP}}(\mathbf{r})|}$ to quantify the spatially resolved optical asymmetry (**Figure S3C**). Consequently, we found that circularly polarized illumination generates significant chiroptical local near fields in the Au nanopattern.

We then simulated the local ROA response of the Au nanopattern based on $\Delta|E(\mathbf{r})|^4$ for the reconstructed 3D models (**Figure 2**). The simulation results revealed pronounced non-zero ROA signals concentrated near the 'edge-dislocation-like' defect cores in the fingerprint-like structures. Notably, compared with the differential near-field map (Figure S3B), the simulated ROA contrast highlights these chiroptical hotspots much more prominently relative to the surrounding regions. This enhancement arises because the simulated ROA scales with the difference between the fourth powers of the local electric-field amplitudes under RCP and LCP excitation, thereby strongly weighting regions of intense near-field concentration. In this way, the stochastic structural asymmetry introduced by the BCP self-assembly synergizes with the electromagnetic response of the plasmonic nanopattern to generate pronounced chiroptical ROA hotspots. This mechanism can yield a non-zero ROA contrast even when probing an achiral substrate vibration, because the observed asymmetry originates from the chiral electric near field rather than from intrinsic vibrational chirality of the substrate.[36,37]

To quantify the optical asymmetry of the ROA response within these hotspots, we further calculated the normalized ROA dissymmetry factor, $\mathrm{CID}(\mathbf{r})$, from the simulated ROA contrast using the following expression (**Figure S3D**).

$$\mathrm{CID}(\mathbf{r}) = \frac{I_{\mathrm{RCP}}(\mathbf{r}) - I_{\mathrm{LCP}}(\mathbf{r})}{I_{\mathrm{RCP}}(\mathbf{r}) + I_{\mathrm{LCP}}(\mathbf{r})} = \frac{|E_{\mathrm{RCP}}(\mathbf{r})|^4 - |E_{\mathrm{LCP}}(\mathbf{r})|^4}{|E_{\mathrm{RCP}}(\mathbf{r})|^4 + |E_{\mathrm{LCP}}(\mathbf{r})|^4} \quad (2)$$

The resulting CID map exhibited clear non-zero signals across the fingerprint-like structures, consistent with the simulated ROA contrast. Importantly, regions identified as chiroptical hotspots in the ROA simulations displayed markedly larger CID values than the surrounding areas, with magnitudes approaching unity, which represents the theoretical maximum. This indicates that these hotspots do not merely produce strong ROA signals but also exhibit pronounced optical asymmetry in the handedness-dependent near-field enhancement. These regions are therefore expected to disproportionately contribute to the measured ROA signal, as the ROA contrast is weighted toward locations that simultaneously exhibit strong near-field enhancement and large handedness-dependent asymmetry.[38-40]

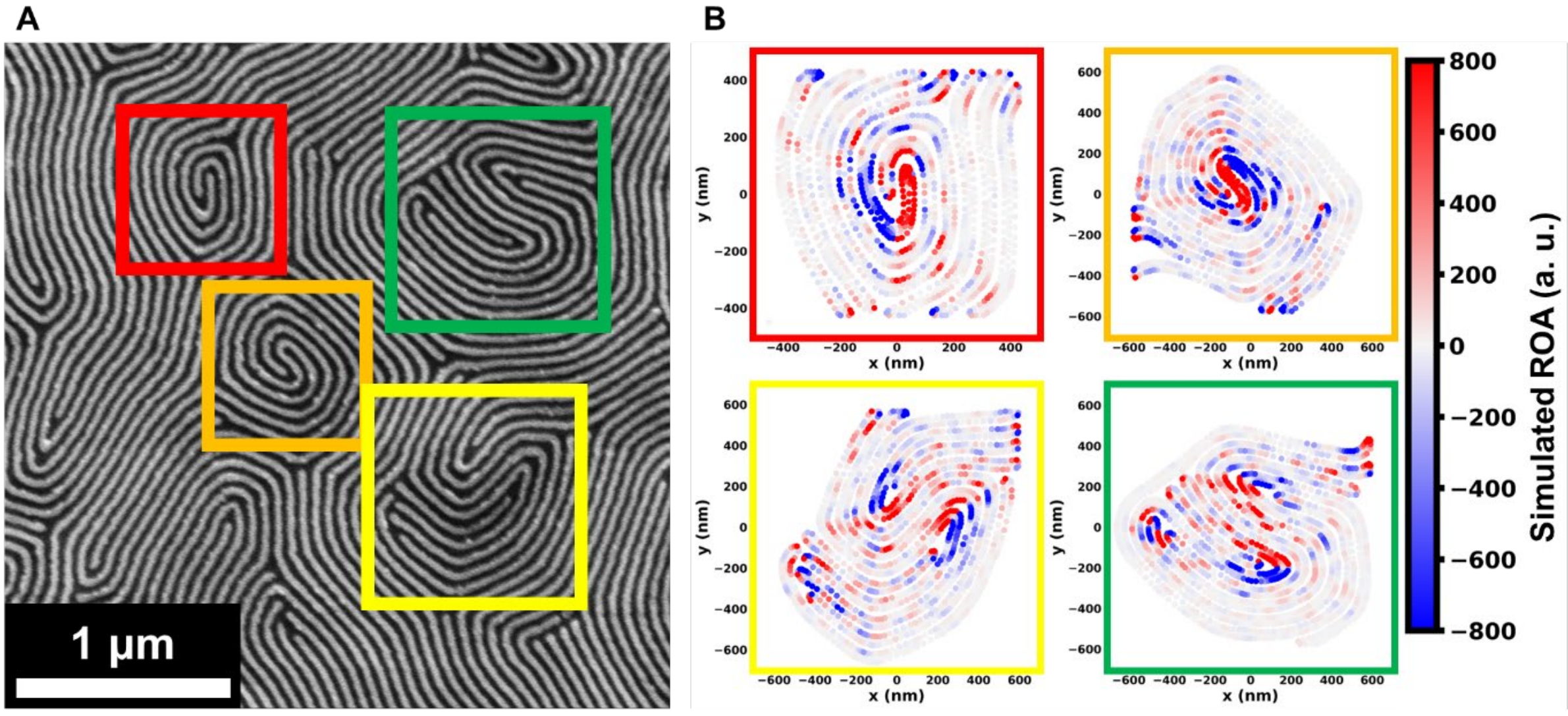

**Figure 2.** Real-space simulation of chiroptical responses in fabricated Au nanopatterns. (A) Top-down SEM image of the fabricated Au nanopatterns, which served as the structural basis for the realistic 3D models used in the ROA simulations. (B) Simulated ROA signals on the reconstructed 3D Au nanopattern models under right- and left-circularly polarized excitation at a wavelength of 633 nm. The colored frames correspond to the specific regions of interest marked in (A).

To validate this result experimentally, we performed large-scale ROA mapping on the fabricated sample. Since the entropy harvesting leverages the morphological stochasticity of the nanopatterns as an entropy source, we collected data from 10,000 discrete spots (100 × 100 grid) in a single Au nanopattern sample. A 633 nm excitation light was employed to match the simulation conditions. To ensure statistical independence in the morphological randomness, the step size between adjacent measurement spots was set to 10 μm, a distance significantly larger than the characteristic domain size observed in the SEM images.

Averaging the Raman spectra from all 10,000 spots revealed a distinct peak at 521 $cm^{-1}$, corresponding to the first-order optical phonon mode of the Si substrate (**Figure S4A**).[41,42] In contrast, the macroscopic average of the ROA spectra across the same spots showed no discernible signal (**Figure S4B**), indicating that the Au nanopatterns possess no dominant handedness on a global scale. However, examining the statistical distribution of the ROA intensity at 521 $cm^{-1}$ for individual spots revealed a different picture. As shown in **Figure 3A**, the local ROA values were not uniformly zero but followed a Gaussian distribution centered near zero (μ = −0.002) with a standard deviation (σ) of 5.394. This suggests that while the sample is globally achiral due to the cancellation of randomly oriented local chiral features, it possesses distinct local handedness at the sub-micrometer scale.

We further classified the 10,000 spots into three groups based on the empirical distribution of ROA intensities (μ = −0.002, σ = 5.394). Spots with ROA values lower than $\mu - z^{*}\sigma$ were labeled as 'Negative', those above $\mu + z^{*}\sigma$ as 'Positive', and the remainder as 'Neutral', where $z^{*} = \Phi^{-1}(\frac{2}{3}) \approx 0.43$ is a fixed Gaussian-tercile constant. The averaged ROA spectra for each group showed that the 'Positive' and 'Negative' groups exhibit strong, opposing ROA signals at 521 $cm^{-1}$, while the 'Neutral' group displayed a negligible response (**Figure 3B**). Crucially, the signal intensities at 521 $cm^{-1}$ for the 'Positive' and 'Negative' groups were dominant compared to the noise floor at other wavenumbers, confirming that the variations in individual spots arose from genuine local chirality rather than random measurement noise.

Subsequently, we investigated the spatial correlation of the measurements to determine whether the ROA signal at a given spot is influenced by its neighbors. Specifically, for the 10,000 measured spots, we analyzed the joint distributions of ROA values between a specific spot and its immediate right-neighbor (**Figure S5A**) and upper-neighbor (**Figure S5B**). The resulting transition count maps reveal that the data points cluster symmetrically around the origin, regardless of the scanning direction. This isotropic clustering indicates that there is no correlation between adjacent measurements, implying that the ROA value at any arbitrary spot is statistically independent of its immediate neighbors. To further quantify this independence over longer distances, we calculated the auto correlation function (ACF) of the ROA sequence as a function of spatial lag (spot interval) across the 10,000 spots (**Figure S5C**). The mean ACF value was calculated to be −0.0001 with a standard deviation of 0.0095. Both values are negligible and effectively zero, conclusively demonstrating that the ROA signals are spatially uncorrelated and statistically independent, regardless of the distance among measurement spots.

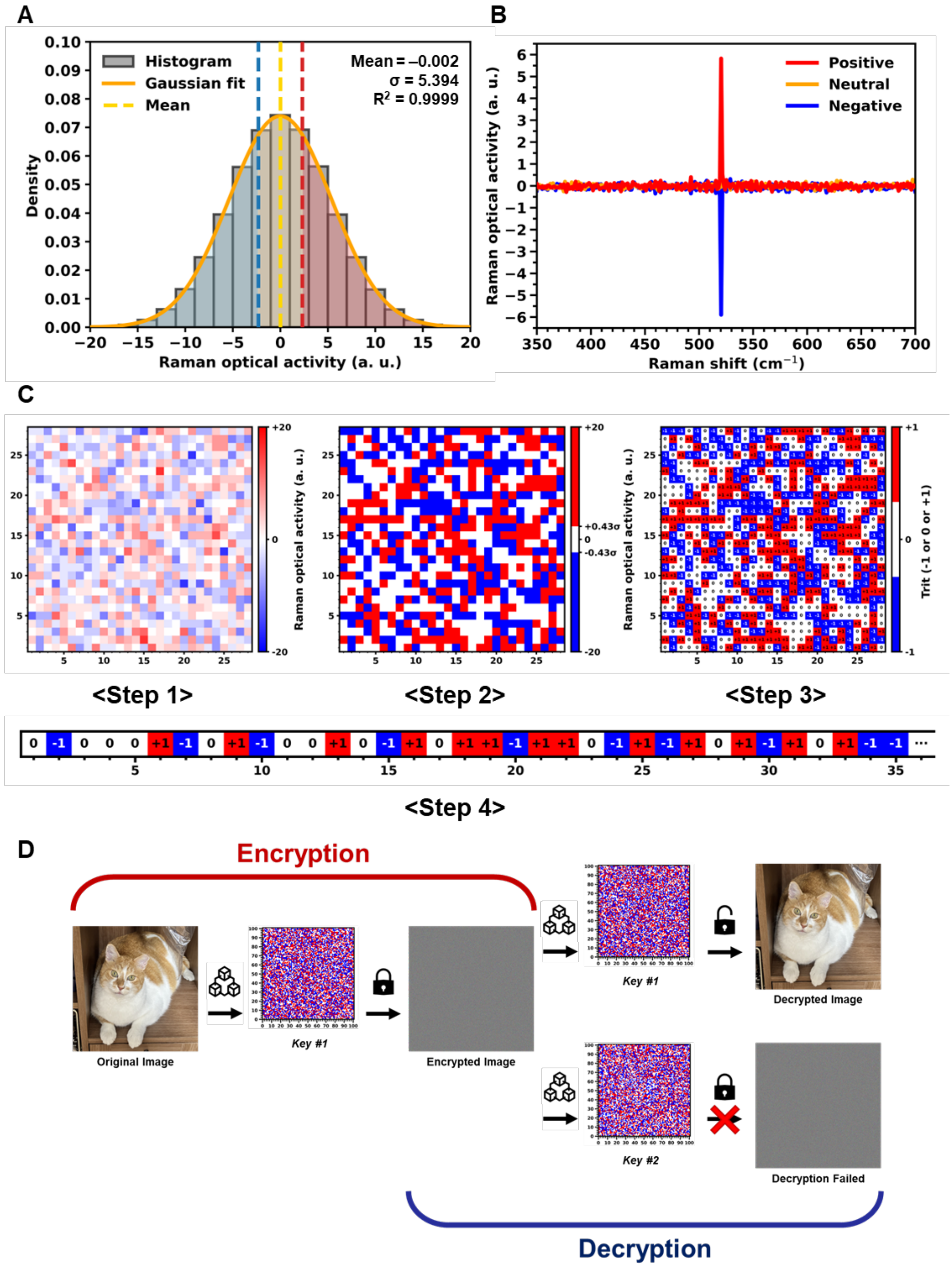

**Figure 3.** Ternary entropy harvesting utilizing local chiroptical responses of Au nanopatterns. (A) Statistical distribution of local chiroptical Raman contrast extracted at the 521 $cm^{-1}$ Raman band, where the handedness-dependent modulation is induced by the stochastic Au nanopatterns. The dashed lines indicate specific thresholds: blue ($\mu-0.43\sigma$), yellow ($\mu$), and red ($\mu+0.43\sigma$). (B) Averaged ROA spectra for the three groups (Positive, Neutral, and Negative) classified based on the $\mu\pm0.43\sigma$ criteria. (C) Stepwise procedure for extracting ternary random sequences from the ROA signals. Note that while a full 100 × 100 grid was employed for the actual sequence generation, a magnified 28 × 28 sub-region is displayed here for visual clarity. (D) Demonstration of image encryption and decryption protocols using the ternary sequences.

We then defined a 100 × 100 grid with a pixel pitch of 10 μm on an arbitrary region of the Au nanopattern and constructed a ROA map. Given that the ROA intensities follow a strong Gaussian distribution (Figure 3A), we applied the classification thresholds established in the previous section to categorize each pixel into 'Positive', 'Neutral', or 'Negative' groups. Subsequently, ternary digits (trits) were assigned to each pixel: '+1' for the 'Positive' group, '0' for 'Neutral', and '−1' for 'Negative'. The resulting 2D ternary map was then flattened row-by-row to generate a ternary random sequence comprising 10,000 trits (**Figure 3C**). These trits were extracted inline from the ROA readout without any external randomness injection.

Prior to a comprehensive statistical analysis of the full sequence, we performed a Monte Carlo simulation to estimate π on the 28 × 28 sub-region shown in Figure 3C to intuitively validate its spatial randomness (**Figure S6**). The simulation revealed that, regardless of the specific trit value (+1, 0, or −1), the estimated π converged to the theoretical constant with a relative error of less than 0.4%. Furthermore, the standard error (*SE*) across the estimates derived from each trit was remarkably low ($SE \approx 0.0056$). We also extracted 10,000 trits from each of three distinct Au nanopattern samples and examined their spatial distributions (**Figure S7**). While all samples consistently exhibited nearly equal trit probabilities (~33%), the resulting heatmaps and sequences displayed distinct, non-repeating patterns. This confirmed that the random sequences generated from different physical samples were distinct and uncorrelated, reflecting the inherent stochasticity of the fabrication process.

To visually demonstrate the cryptographic quality of the generated random sequences and the statistical independence between different physical samples, we performed image encryption using an industry-standard cryptographic protocol (AES-256) (**Figure 3D**). Specifically, a high-resolution digital photograph (original image) was selected as the plaintext. The 10,000-trit sequence extracted from the Au nanopattern was processed via a standardized secure hash function (SHA-256) to derive the encryption key. The resulting encrypted image manifested as a noise-like pattern with high entropy, effectively masking all morphological features of the original image. When the ciphertext was decrypted using the key derived from the identical ternary sequence (Key #1), the original image was successfully reconstructed with pixel-perfect accuracy. In stark contrast, an attempt to decrypt the image using a key derived from a ternary sequence extracted from a different Au nanopattern (Key #2) resulted in a complete decryption failure, yielding only random noise.

We next expanded the dataset by employing 64 distinct Au nanopattern samples fabricated under identical conditions. From each sample, we extracted 10,000 trits using a 100 × 100 grid with a 10 μm pitch, resulting in a total dataset of 640,000 trits for comprehensive statistical validation. First, we analyzed the occurrence probability of each trit value across the entire dataset (**Figure S8A**). The measured probabilities were 33.36% for '−1', 33.36% for '0', and 33.28% for '+1'. These values closely converge to the theoretical expectation of approximately 33.33% for an ideal equiprobable ternary distribution. This demonstrates that the entropy harvester operates in a highly balanced manner without exhibiting bias toward any specific state. Furthermore, to quantitatively assess the unpredictability of the generated sequences, we calculated the Shannon entropy (**Figure S8B**).[43] The resulting entropy was determined to be 1.58496 bits per trit, which is virtually identical to the theoretical maximum for a ternary system ($\log_2 3$ bits per trit). This confirms that the generated trits possess high-purity randomness with negligible statistical bias.

To rigorously verify the physical origin of this high-purity randomness, we conducted a control experiment on a bare Si wafer devoid of Au nanopatterns, applying the identical ROA measurement and trit extraction protocols (**Figure S9**). In stark contrast to the Au nanopattern samples, the control group exhibited a severely biased distribution: the occurrence probabilities were 10.94% for '−1', 79.30% for '0', and 9.76% for '+1'. Consequently, the Shannon entropy plummeted to 0.942 bits per trit, falling significantly short of the theoretical maximum. This degradation is directly attributed to the absence of locally chiral Au nanopatterns, which led to a substantial decrease in non-zero ROA signals. These results definitively confirm that the stochastic local chirality inherent to the Au nanopatterns serves as the fundamental entropy source for the high-quality ternary entropy harvesting.

A crucial advantage of our ternary entropy harvesting system, highlighted by the measured Shannon entropy, is its superior information density compared to binary systems. While the theoretical entropy limit for conventional binary entropy harvesting is 1 bit per digit, our chiroptical ternary system achieves ~1.585 bits per digit. From an information-theoretic perspective, this implies that the proposed ternary system can accommodate approximately 58.5% more information per digit than a binary counterpart, offering significantly enhanced data capacity.

While the equiprobable distribution of trits and a Shannon entropy value virtually identical to the theoretical maximum are critical indicators, they constitute merely necessary conditions, not sufficient criteria for an entropy harvester.[44] Global equiprobability does not preclude the existence of inter-trit correlations; if the generated sequence exhibits predictable patterns or dependencies, it loses its stochastic nature and fails to function as an entropy-harvesting system.[45] To rigorously verify the statistical independence of the generated trits, we calculated the ACF for a sequence of 10,000 trits extracted from a single Au nanopattern (**Figure S10A**). The analysis yielded a mean ACF value of −0.0002 with a standard deviation of 0.0095. Both values are negligible and approach zero, indicating a complete absence of statistical correlation within the 1D sequence.

Beyond the internal independence within a single sequence, the absence of correlation between different ternary sequences is equally critical for an entropy harvesting system. Even if a single sequence appears statistically random, any correlation between sequences derived from different sources would undermine the unpredictability essential for cryptographic security.[44,46,47] To verify this, we calculated the inter-sample correlation coefficients between 64 distinct 1D trit sequences extracted from separate Au nanopattern samples (**Figure S10B**). Consistent with the previously discussed ACF results, the correlation coefficients exhibited a mean value of 0.0003 and a standard deviation of 0.0098. These values, both remarkably close to zero, provide strong statistical evidence that there is no linear dependency or shared pattern between sequences. This confirms that each ternary sequence generated from our chiroptical system is statistically distinct and uncorrelated with others, reflecting a high degree of stochasticity provided by the independent physical entropy sources.

Furthermore, to ensure that this independence extends to the spatial domain, we analyzed the transition probabilities between adjacent pixels in the 100 × 100 ternary heatmap. We calculated the occurrence probabilities of trits adjacent to each specific trit value in both orthogonal and diagonal directions (**Figure S10C, D**). The results demonstrated uniform transition probabilities across all directions, confirming that our ternary entropy harvester exhibits no discernible correlations in either 1D temporal sequences or 2D spatial arrangements.

While the ACF and inter-sample correlation coefficient analyses provide strong evidence for the absence of intra-sequence correlations and inter-sequence linear dependencies, these metrics alone do not fully quantify the performance of an entropy harvesting system. A near-zero correlation coefficient does not necessarily preclude the possibility that two discrete sequences share an unusually high fraction of identical symbols. Therefore, we additionally measured the symbol-wise dissimilarity between ternary outputs using the normalized Hamming distance, defined as the fraction of trit positions at which two sequences differ.

Specifically, we measured the normalized Hamming distance for all pairwise combinations of 64 ternary sequences (10,000 trits per sequence), yielding a total of 2016 comparisons (**Figure S10E**). The normalized Hamming distance was used here as a direct measure of trit-level dissimilarity, i.e., the fraction of positions at which two sequences differ. The resulting distribution exhibited a mean of 0.6665 with a standard deviation of 0.0048 (**Figure S10F**). Notably, the mean is essentially indistinguishable from the ideal value of 2/3 expected for two independent, equiprobable ternary sequences, and the narrow spread indicates highly consistent behavior across the full set of pairwise comparisons.

In addition, to verify that spatially separated regions within a single 2D trit map remain statistically independent at the trit level, we partitioned one 100 × 100 trit map into 25 non-overlapping blocks of 20 × 20 trits and measured the blockwise inter-Hamming distances among these blocks (**Figure S10G**). The resulting distribution exhibited a mean of 0.6666 with a standard deviation of 0.0235 (**Figure S10H**), again closely matching the ideal value of 2/3. The broader spread is consistent with the shorter sequence length per block (400 trits), which naturally increases statistical fluctuations. Together, these results highlight the high quality of the ternary output and experimentally demonstrate that ternary sequences obtained from distinct Au nanopattern samples, as well as spatially separated regions within a single map, maintain robust randomness and statistical independence.

While the metrics above (symbol balance, entropy, ACF, correlation analysis, and Hamming-distance statistics) collectively indicate minimal bias and negligible correlations in our ternary outputs harvested from inherent structural stochasticity, these analyses alone do not provide a fully comprehensive randomness assessment. We therefore benchmarked the performance of our chiroptical ternary entropy harvester against an established statistical standard by applying the NIST SP 800-22 test suite published by the National Institute of Standards and Technology (NIST). Since the suite is defined for binary inputs, the ternary sequences were deterministically expressed as a binary bitstream via arithmetic coding, without applying any whitening or post-processing. It should be noted that this step does not introduce additional randomness or modify the ternary sequences, as deterministic transformations cannot increase the Shannon entropy of the source.[48,49] Accordingly, the 1,014,450-bit dataset derived from 640,000 trits passed all applicable NIST SP 800-22 tests (**Table 1**). Given the breadth of this suite, which was designed to detect diverse forms of non-randomness beyond simple bias or linear correlation, these outcomes provide stringent statistical validation of our chiroptical ternary entropy harvester and support its suitability as a high-quality physical entropy source for practical use.

To complement conventional statistical tests, we further evaluated the predictability of the ternary output using machine-learning-based sequence modeling. The rationale is straightforward: if a model can predict upcoming trits from preceding trits with accuracy above chance, the sequence contains learnable structure. For ternary symbols, random guessing yields an expected accuracy of 1/3 (33.3%), and the corresponding baseline for the per-trit categorical cross-entropy is $\ln 3$.

We first applied a Markov predictor to probe short-range dependencies by varying the Markov order as well as the conditional and prediction lengths (**Figure 4A, B and Figure S11**). Across all conditions, the prediction accuracy remained close to 33.3% and the minimum validation cross-entropy over training epochs did not decrease below $\ln 3$ within statistical uncertainty, indicating the absence of learnable local rules. We further challenged the sequence using a gated recurrent unit (GRU) model, which can capture longer-range and more complex dependencies, by varying the lag and the amount of training data (**Figure 4C, D and Figure S12–S19)**. Under all tested conditions, the accuracy again stayed near the chance level and the minimum validation cross-entropy did not fall below $\ln 3$ within statistical uncertainty, suggesting that the generated ternary sequences do not exhibit readily learnable long-range context such as repeating patterns or higher-order dependencies.

In addition, we employed a variational autoencoder (VAE)-based generator to test whether the ternary outputs contain detectable latent structure. In this analysis, a segment of the ternary sequence was provided as a conditional input, and the model was tasked with predicting and generating subsequent trits (**Figure 4E, F and Figure S20–S23**). Across a range of conditional and generation lengths, the accuracy remained near 33.3% and the minimum validation cross-entropy likewise did not decrease below $\ln 3$ within statistical uncertainty. Taken together, these results provide further evidence that the chiroptical ternary entropy harvester outputs lack readily exploitable structure and are not meaningfully predictable by the tested machine-learning models.

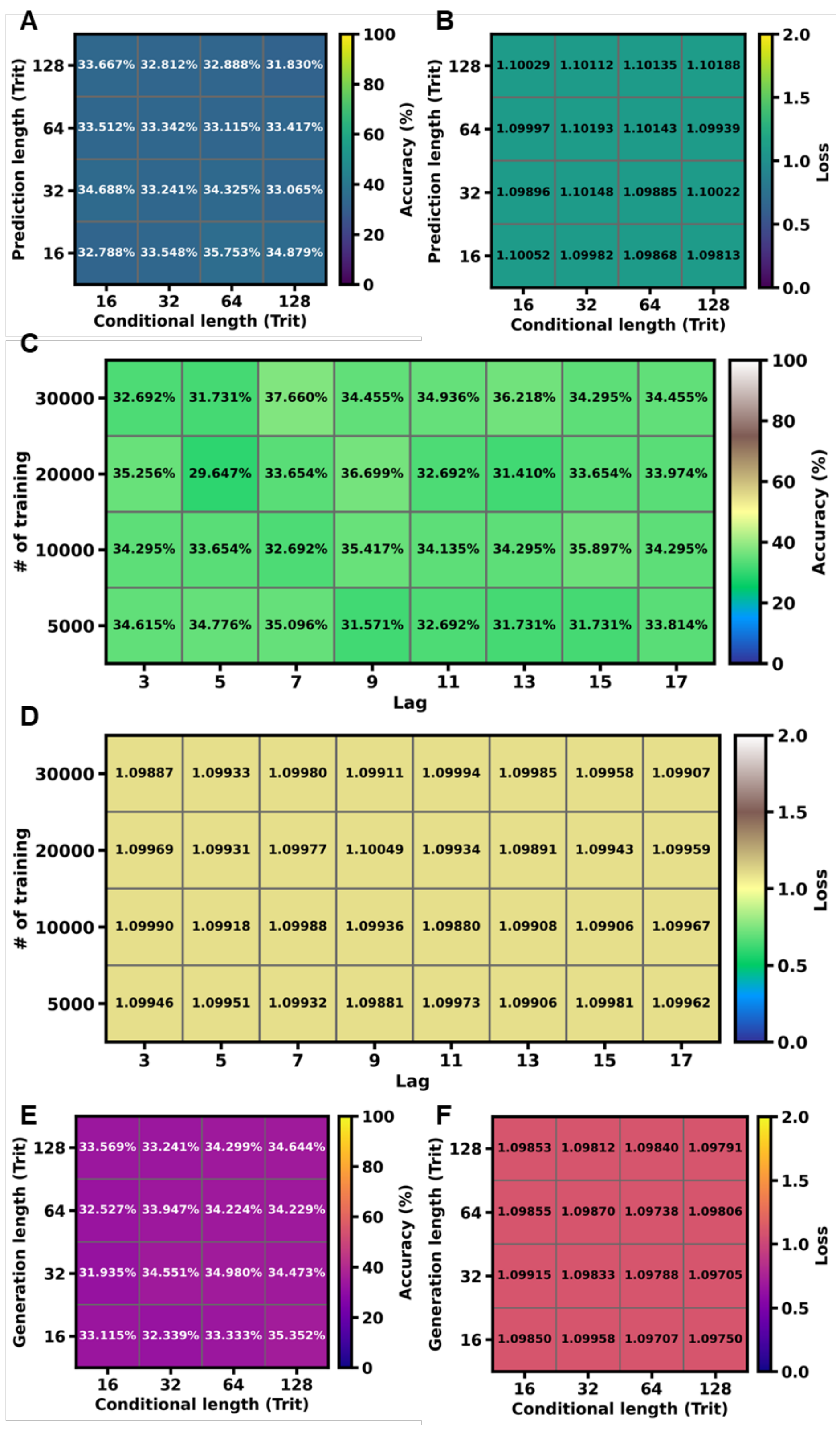

**Figure 4.** Machine-learning–based predictability analysis of the chiroptical ternary entropy harvester output. Markov prediction results showing (A) validation accuracy and (B) per-trit categorical cross-entropy (loss) as a function of conditional length and prediction length. GRU-based sequence prediction results showing (C) validation accuracy and (D) minimum validation cross-entropy across training epochs as a function of lag and training-set size. VAE-based conditional generation results showing (E) validation accuracy and (F) minimum validation cross-entropy across training epochs as a function of conditional length and generation length.

While the preceding analyses establish the statistical quality of our chiroptical ternary entropy harvester, its translation to real-world applications necessitates exceptional physical stability. For a physically encoded entropy material, this robustness is not a secondary property but a strict prerequisite to preserve the stored optical alphabet against environmental perturbations.[50,51] We therefore subjected the Au nanopatterns to thermal, cryogenic, mechanical and acidic stresses and found that both the fingerprint-like morphology and the ternary output statistics remained intact.

Specifically, we subjected identically fabricated Au nanopattern-on-Si wafers to harsh environmental stressors, including heating at 150 °C (oven), cryogenic exposure in liquid nitrogen (−196 °C), ultrasonication in water, and immersion in 1 M HCl for 1 h. After each treatment, the nanopatterns were examined by SEM, and ternary heatmaps and trit sequences were extracted using the same acquisition and digitization workflow. We then assessed symbol balance and the inter-sample normalized Hamming distance as primary indicators of output stability (**Figure 5 and Figure S24**). Representative SEM images confirmed that the characteristic fingerprint-like nanostructure was preserved regardless of the applied harsh conditions. Consistently, the extracted ternary outputs maintained near-equiprobable trit populations (approximately 33.3% per symbol), and the inter-sample normalized Hamming distance remained close to the ideal value of 2/3 expected for independent, equiprobable ternary sequences. These results show that the Au nanopattern platform retains its structural integrity under the tested stresses and that the statistical characteristics of the extracted ternary entropy harvester outputs remain stable even after exposure to severe physical and chemical environments.

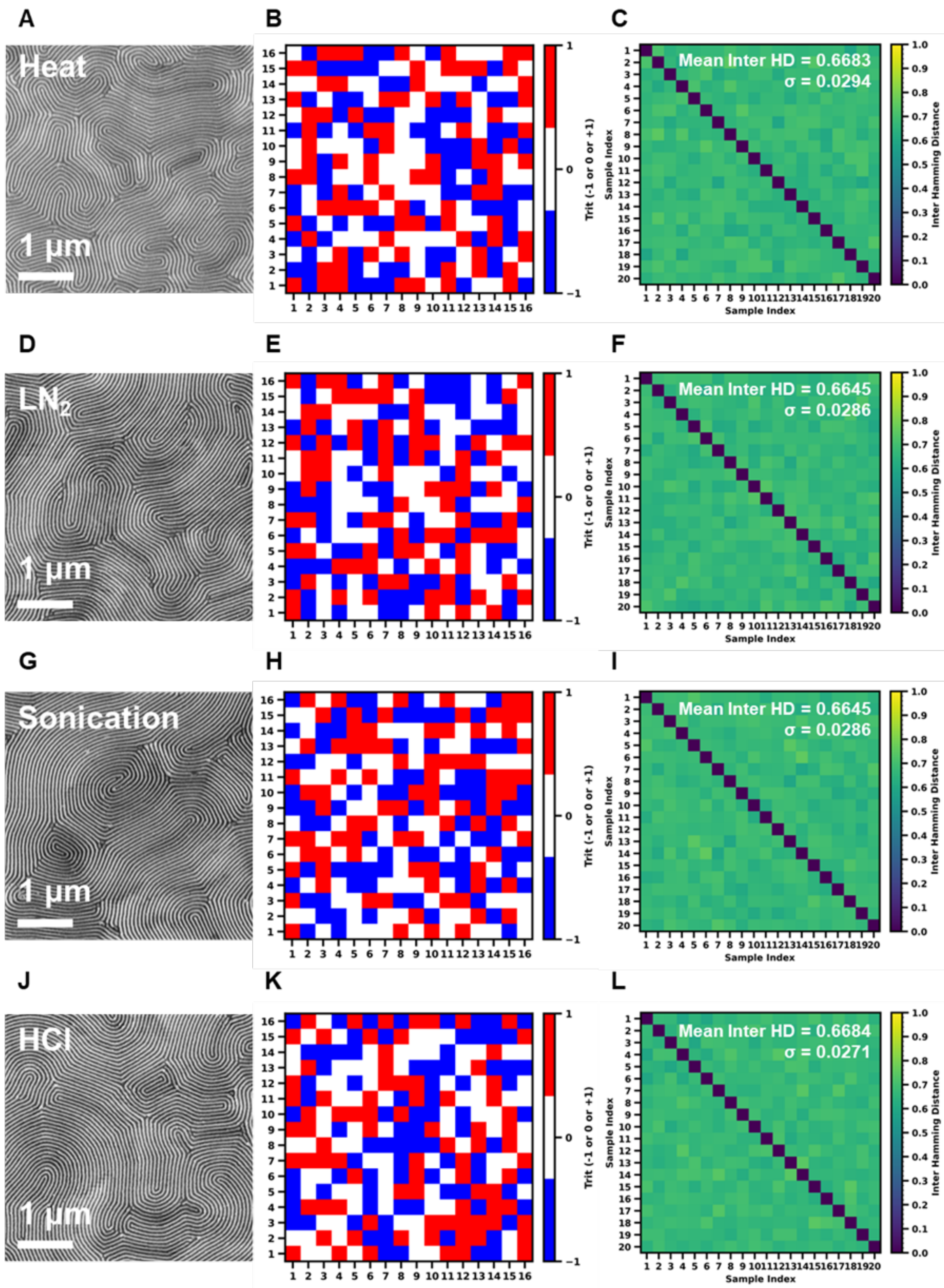

**Figure 5.** Robustness assessment of the Au nanopattern-on-Si platform for ternary entropy harvester operation under harsh environments. After heating at 150 °C: (A) SEM image of the Au nanopattern, (B) extracted ternary heatmap, and (C) inter-sample normalized Hamming-distance matrix used to evaluate output stability. After cryogenic exposure in liquid nitrogen (−196 °C): (D) SEM image, (E) ternary heatmap, and (F) inter-sample normalized Hamming-distance matrix. After ultrasonication in water: (G) SEM image, (H) ternary heatmap, and (I) inter-sample normalized Hamming-distance matrix. After immersion in 1 M HCl for 1 h: (J) SEM image, (K) ternary heatmap, and (L) inter-sample normalized Hamming-distance matrix.

Beyond demonstrating robustness under harsh environments, it is also important to verify that the chiroptical entropy harvesting is not restricted to a single nanopattern geometry. As shown in Figure S1, the center-to-center spacing and the line width of the Au nanopatterns can be systematically tuned by adjusting the molecular weight of BCPs. We therefore fabricated a series of Au nanopattern-on-Si samples with different line widths and applied the same ROA-based digitization framework to extract ternary outputs (**Figure S25**). Across these width-tuned samples, the trit occupancy remained near equiprobable (approximately 33.3% per symbol), and the inter-sample normalized Hamming distance stayed close to the ideal value of 2/3 expected for independent, equiprobable ternary sequences. Together, these results show that the proposed chiroptical ternary entropy harvesting system is compatible with geometry-tunable nanopattern fabrication and that its statistical performance is not contingent on a single optimized structure.

While the preceding results demonstrate the statistical quality and robustness of our chiroptical ternary entropy harvesting using Si-supported Au nanopatterns, practical integration further requires compatibility with diverse functional substrates and scalable fabrication. Extending the platform beyond Si is valuable because many target applications demand substrates that provide additional functions, such as mechanical flexibility for wearable electronics, electrical conductivity for electrode-integrated devices, or optical transparency for optoelectronic interfaces. We therefore fabricated Au nanopatterns on representative platforms spanning flexible polymers (polydimethylsiloxane; PDMS), conductive electrodes (glassy carbon), and transparent electrodes (indium tin oxide; ITO) (**Figure S26A–L and Figure S27A–C**). For each substrate, ROA acquisition and ternarization were performed by probing a dominant Raman band intrinsic to the substrate, specifically 2910 $cm^{-1}$ for PDMS (C–H stretching), 1330 $cm^{-1}$ for glassy carbon (D band), and 475 $cm^{-1}$ for ITO (lattice mode).[52-54] Across all substrates, the extracted ternary outputs exhibited near-equiprobable trit occupancies (~33% per symbol), a Shannon entropy close to the ideal value for ternary sequences (~1.585 bits per trit), and inter-sample normalized Hamming distances close to 2/3, as expected for independent, equiprobable ternary sequences. Overall, our chiroptical ternary entropy harvesting approach can be readily adapted to a broad range of functional substrates while maintaining high-quality ternary randomness, supporting its potential for integration into diverse device architectures.

For practical deployment, a physical entropy harvesting system must be compatible not only with diverse substrates but also with scalable manufacturing. Wafer-scale patterning enables high-throughput fabrication and cost-effective integration into established device-processing workflows. To assess scalability, we extended our fabrication from small Si chips to a full 2-inch Si wafer patterned with Au nanopatterns and subsequently extracted ternary outputs using the same ROA-based digitization protocol (**Figure S26M–P and Figure S27D**). Notably, the wafer-scale samples produced ternary outputs with near-equiprobable trit occupancies, a Shannon entropy close to the ternary ideal, and inter-sample normalized Hamming distances near the ideal value expected for independent, equiprobable ternary sequences.

## Conclusions

We have demonstrated that self-assembled nanoscale disorder can be transformed from an unavoidable imperfection in nanofabrication into a functional source of physical entropy. By replicating block copolymer-derived fingerprint-like morphologies into plasmonic Au nanopatterns, we create locally heterogeneous chiroptical landscapes whose handedness-dependent optical responses can be directly harvested through ROA mapping. Because this response is intrinsically signed, the material output is naturally digitized into ternary symbols rather than forced into a binary format.

The resulting ternary sequences exhibit near-ideal symbol balance, negligible spatial and inter-sample correlations, Shannon entropy approaching the theoretical ternary limit, and no detectable predictability under statistical or machine-learning-based analyses. The platform further retains its entropy-harvesting performance across environmental stresses, tunable nanopattern geometries, diverse functional substrates and wafer-scale fabrication. This work opens a route towards compact cryptographic keys, tamper-resistant hardware identifiers and entropy-rich security layers for emerging edge, wearable and distributed electronics.

**Tables**

**Table 1.** NIST SP 800-22 statistical test suite applied to the binary bitstream (1,014,450 bits) deterministically encoded from the ternary entropy harvester output (640,000 trits).

| Test | p-value | Result | Remarks |
|---|---|---|---|
| **Frequency** | 0.530343 | PASS | |
| **Block Frequency (M = 128)** | 0.575835 | PASS | |
| **Cumulative Sums (Forward)** | 0.387253 | PASS | |
| **Cumulative Sums (Reverse)** | 0.904579 | PASS | |
| **Runs** | 0.202234 | PASS | |
| **Longest Runs of Ones** | 0.260066 | PASS | |
| **Rank** | 0.672544 | PASS | |
| **Discrete Fourier Transform** | 0.324001 | PASS | |
| **Non-overlapping Template (m = 9)** | 0.015115 | PASS | Smallest of the multiple p-values. m = The length in bits of the template |
| **Overlapping Template (m = 9)** | 0.645786 | PASS | m = The length in bits of the template |
| **Universal Statistical** | 0.567660 | PASS | |
| **Approximate Entropy (m = 10)** | 0.846558 | PASS | m = The length of each block |
| **Serial (m = 16)** | 0.262396 | PASS | p-value$_1$, m = The length of each block |
| | 0.428069 | PASS | p-value$_2$, m = The length of each block |
| **Linear Complexity (M = 500)** | 0.612183 | PASS | M = The length of each block |

## Methods

### Materials

Polystyrene-*block*-poly(methyl methacrylate) random copolymer terminated α-hydroxy, ω-TEMPO-moiety (PS-*r*-PMMA) and all polystyrene-*block*-poly(methyl methacrylate) (PS-*b*-PMMA) with different average molecular weights ($M_n$) were purchased from Polymer Source. Toluene (99.8%, anhydrous) and tetrahydrofuran (THF, 99.5%, stabilized) were purchased from Sigma Aldrich. Sylgard 184 base and curing agent were purchased from Omniscience. 1 mm-thick glassy carbon substrates and hydrofluoric acid (HF, 48%) were purchased from Thermo Fisher Scientific. P-type, (100)-oriented 2-inch Si wafers, p-type, (100)-oriented 4-inch Si/$SiO_2$ wafers with a 5,000 Å-thick wet oxide layer, and indium tin oxide (ITO) substrates were purchased from iTASCO.

### Fabrication of stochastic nanopatterns by BCP self-assembly

Prior to BCP self-assembly, the surface energy of the wafers was neutrally modified by using brush layer. The brush solution was prepared by dissolving 1 wt% PS-*r*-PMMA in toluene and spin-coated onto ultraviolet-ozone-treated wafers at 1000 rpm for 60 s. The wafers coated with the brush solution were then annealed in a vacuum oven at 160 °C for 12 h. Three different PS-*b*-PMMA solutions were prepared by blending one of three high-$M_n$ PS-*b*-PMMA with distinct $M_n$ value (49.5 kg $mol^{-1}$, 111 kg $mol^{-1}$, 170 kg $mol^{-1}$ for PS and 48.5 kg $mol^{-1}$, 115 kg $mol^{-1}$, 178 kg $mol^{-1}$ for PMMA, respectively) with a low-$M_n$ PS-*b*-PMMA (6 kg $mol^{-1}$ for PS and 6 kg $mol^{-1}$ for PMMA) in toluene. The blends were prepared at weight ratios of 9:1, 7.5:2.5, and 7:3, respectively, to yield solutions containing 2 wt% polymer in toluene. The BCP solutions were spin-coated onto the wafers at 2500 rpm for 40 s. Phase separation was induced by solvent annealing in THF vapor atmosphere, followed by thermal annealing in a vacuum oven at 250 °C.

### Fabrication of Au nanopatterns

Au nanopatterns were fabricated with Au deposition followed by lift off. Prior to Au deposition, PMMA blocks were selectively removed to form PS trenches via inductively coupled plasma reactive ion etching, using an $O_2$ gas source with a power of 50 W and a bias power of 30 W. An Au layer was deposited onto the entire patterned area using an e-beam evaporator. A thin Cr layer beneath the Au layer was used as an adhesion layer. Subsequently, a lift-off process with sonication was performed.

### Fabrication of Au nanopatterns on various substrates

To demonstrate substrate versatility, Au nanopatterns were transferred onto PDMS, ITO, and glassy carbon substrates. PDMS substrates were prepared by mixing the Sylgard 184 base and curing agent at a ratio of 10:2, pouring the mixture into a plastic dish, and curing in a vacuum oven at 70 °C for 12 h. The substrate surfaces were made hydrophilic by $O_2$-plasma reactive ion etching under a source power of 50 W. ITO and glassy carbon substrates were used without further treatment. Au nanopatterns were initially fabricated on Si/$SiO_2$ donor wafers. The $SiO_2$ layer beneath the Au nanopatterns was subsequently etched with HF vapor, enabling the release of the nanopatterns from the donor wafers. The released Au nanopatterns were then transferred onto the desired receiver substrates.

### Characterization of Au nanopatterns

SEM analysis was performed with an S-4800 (Hitachi). TEM and EDS analysis were performed with a Talos F200X G2 (Thermo Fisher Scientific). Sample preparation for cross-sectional TEM analysis was done by using Helios G4 (FEI). GISAXS analysis was performed at the 3C beamline of the Pohang Accelerator Laboratory (PAL) in Korea. Raman and ROA measurements were performed with a LabRAM HR Evolution (Horiba).

**Supplementary Information**

Additional figures include SEM images, GISAXS patterns, Raman spectroscopy spectra, computational simulation results, detailed statistical metrics, and machine learning results.

**Author Contributions**

W. J., S. J., and K. K. contributed equally to this work. W. J., S. J., and K. K. conceived the study. K. K. fabricated and analyzed Au nanopatterns using SEM and GISAXS. W. J. analyzed Au nanopatterns using TEM and EDS. S. J. performed Raman and ROA measurements. W. J. developed the ROA-based ternary digitization method and carried out statistical analysis. S. J. performed machine-learning-based predictability assessments and interpreted the results. D. L. performed computation simulations. W. J. and J. Y. wrote the manuscript with comments from all authors. J. Y. and S. O. K. supervised the entire research work.

**Conflict of Interest**

The authors declare no conflict of interest.

**Acknowledgements**

W. J., S. J., D. L., and J. Y. were financially supported by the National Research Foundation of Korea (NRF) grant funded by the Korea government (MSIT) (RS-2026-25494716), the KAIST Center for Contemplative Science Internal Research Project, and the Merck-KAIST Research Collaboration (G01260043). S. O. K. and K. K. were financially supported by the InnoCORE program of the Ministry of Science and ICT (N10260097) and the National Research Foundation of Korea (NRF) grant funded by the Korea government (MSIT) (RS-

2025-00521316).

**Data Availability Statement**

The data that support the findings of this study are available in the supplementary information of this article.

# Supplementary Information

## Chiroptical Ternary Entropy Harvesting from Self-Assembled Block Copolymer Nanopatterns

*Woojkin Jung[1,#], Serin Jeong[1,#], Kyulim Kim[1,#], Dongkyu Lee[1], Sang Ouk Kim[1,4,*], and Jihyeon Yeom[1,2,3,4,*].*

[1] Department of Materials Science and Engineering, Korea Advanced Institute of Science and Technology (KAIST), 291 Daehak-ro, Yuseong-gu, Daejeon 34141, Republic of Korea

[2] Department of Biological Sciences, Korea Advanced Institute of Science and Technology (KAIST), 291 Daehak-ro, Yuseong-gu, Daejeon 34141, Republic of Korea

[3] Institute for Health Science and Technology, Korea Advanced Institute of Science and Technology (KAIST), 291 Daehak-ro, Yuseong-gu, Daejeon 34141, Republic of Korea

[4] Institute for the NanoCentury, Korea Advanced Institute of Science and Technology (KAIST), 291 Daehak-ro, Yuseong-gu, Daejeon 34141, Republic of Korea

[*] Corresponding author

[#] Authors contributed equally

## Supplementary Figures

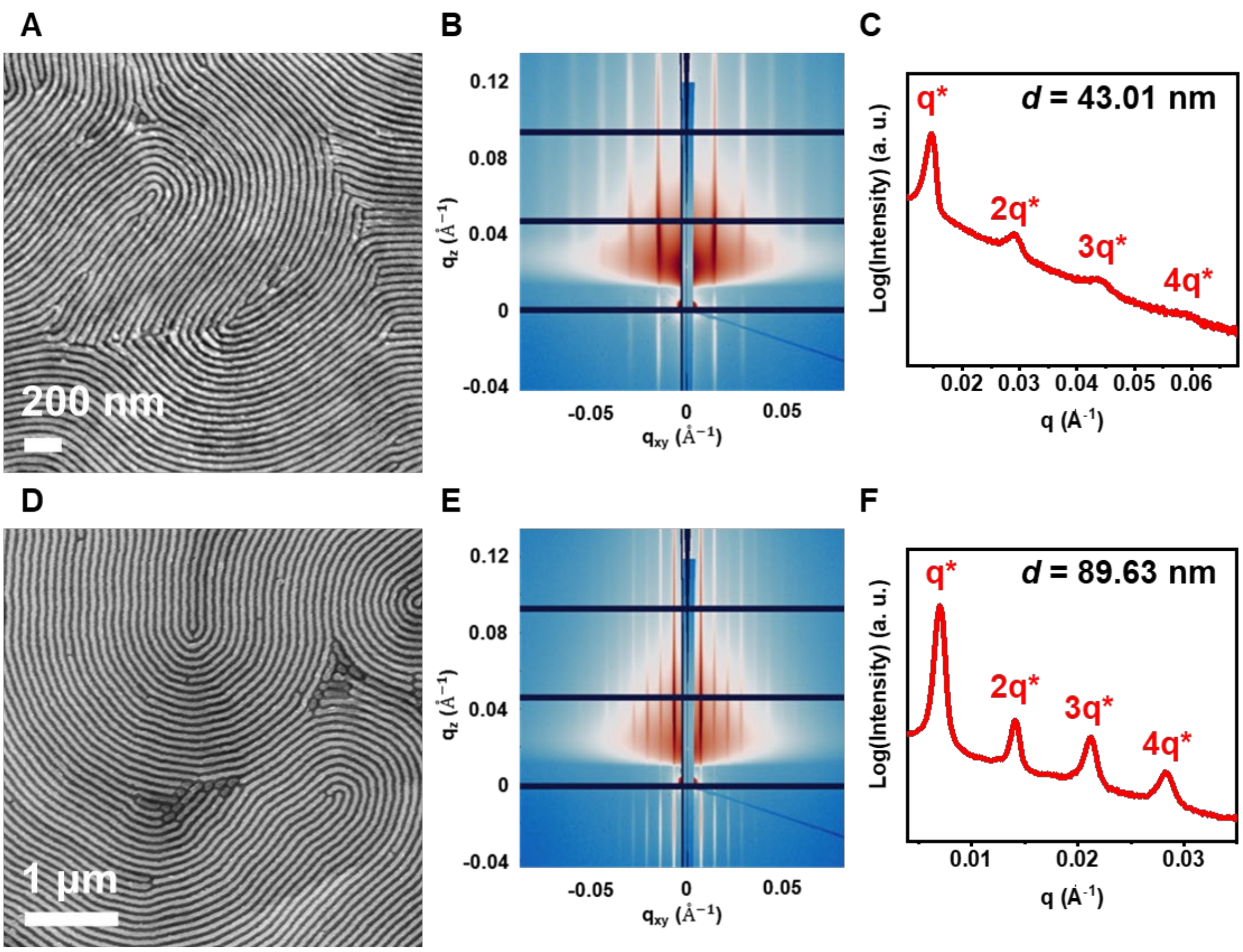

**Figure S1.** Dependence of Au nanopattern structures on the molecular weight of PS-*b*-PMMA. (A–C) Structural analysis of nanopatterns derived from 49.5k-*b*-48.5k PS-*b*-PMMA: (A) Top-down SEM image, (B) 2D GISAXS pattern, and (C) 1D in-plane intensity profile. (D–F) Corresponding analysis for nanopatterns derived from 170k-*b*-178k PS-*b*-PMMA: (D) Top-down SEM image, (E) 2D GISAXS pattern, and (F) 1D in-plane intensity profile.

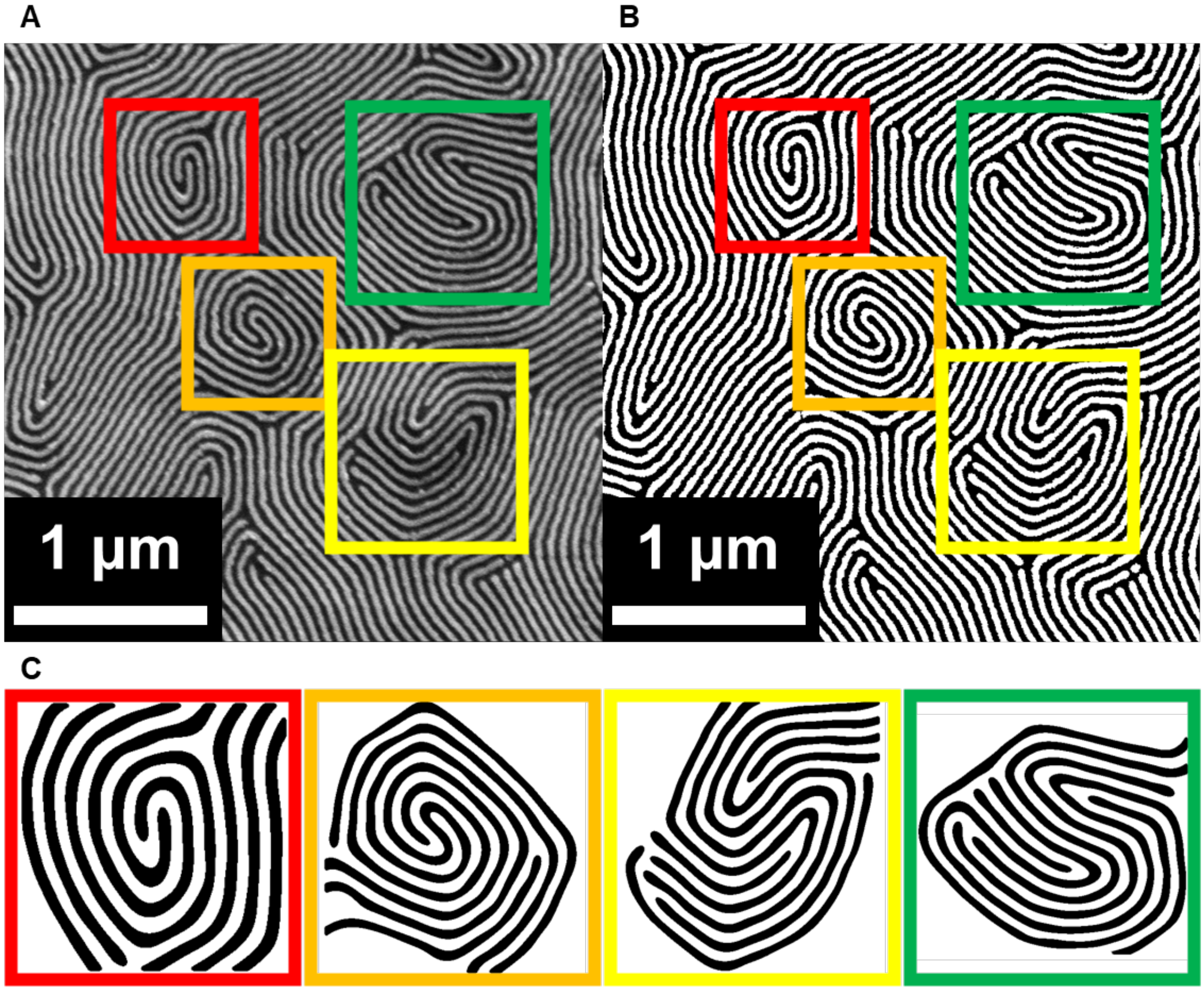

**Figure S2.** Extraction of realistic 3D models for chiroptical FEM simulation. (A) Top-down SEM image of the fabricated Au nanopatterns and (B) the corresponding binarized image. In the binarized image, the bright regions represent the Au nanostructures. (C) Individual nanoscale features extracted from the binarized image. The colored frames correspond to the specific regions of interest marked in (A) and (B). Note that the contrast in (C) is inverted relative to (B); here, the black regions represent the Au patterns. For the 3D FEM simulations, these 2D features were vertically extruded by 11.6 nm, a thickness determined via cross-sectional TEM analysis.

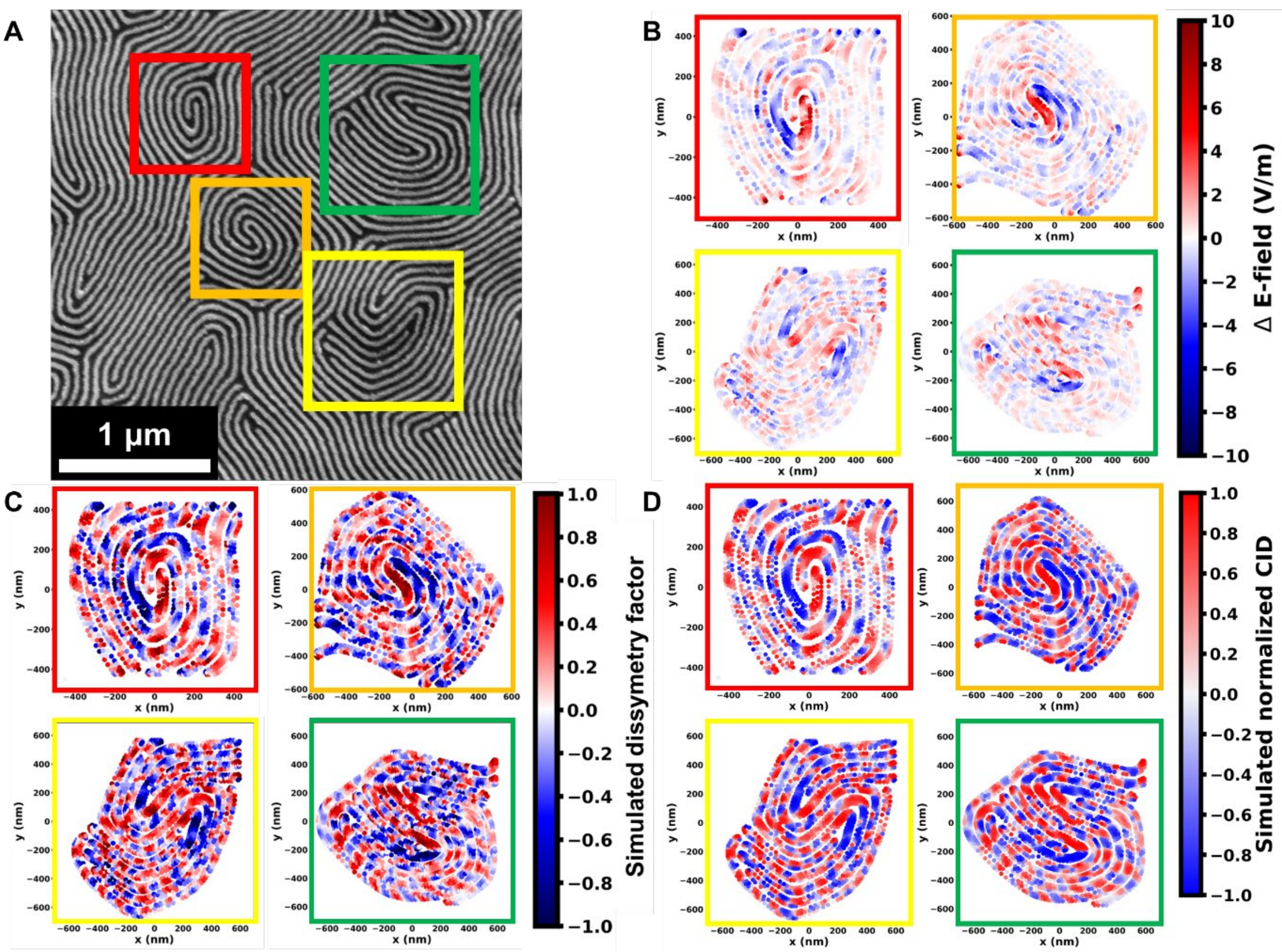

**Figure S3.** Real-space simulation of local chirality and its chiroptical responses in fabricated Au nanopatterns. (A) Top-down SEM image of the fabricated Au nanopatterns, which served as the structural basis for the realistic 3D models used in the simulations. (B) Calculated differential electric field distributions (ΔE) on the reconstructed 3D Au nanopattern models under right- and left-circularly polarized excitation at a wavelength of 633 nm. (C) Spatial maps of the local dissymmetry factor calculated from the normalized RCP/LCP local-field difference, quantifying the near-field optical asymmetry. (D) Spatial maps of the normalized ROA dissymmetry factor (CID) calculated from the simulated ROA signals, quantifying the optical asymmetry of the ROA responses. The colored frames in (B), (C), and (D) correspond to the specific regions of interest marked in (A).

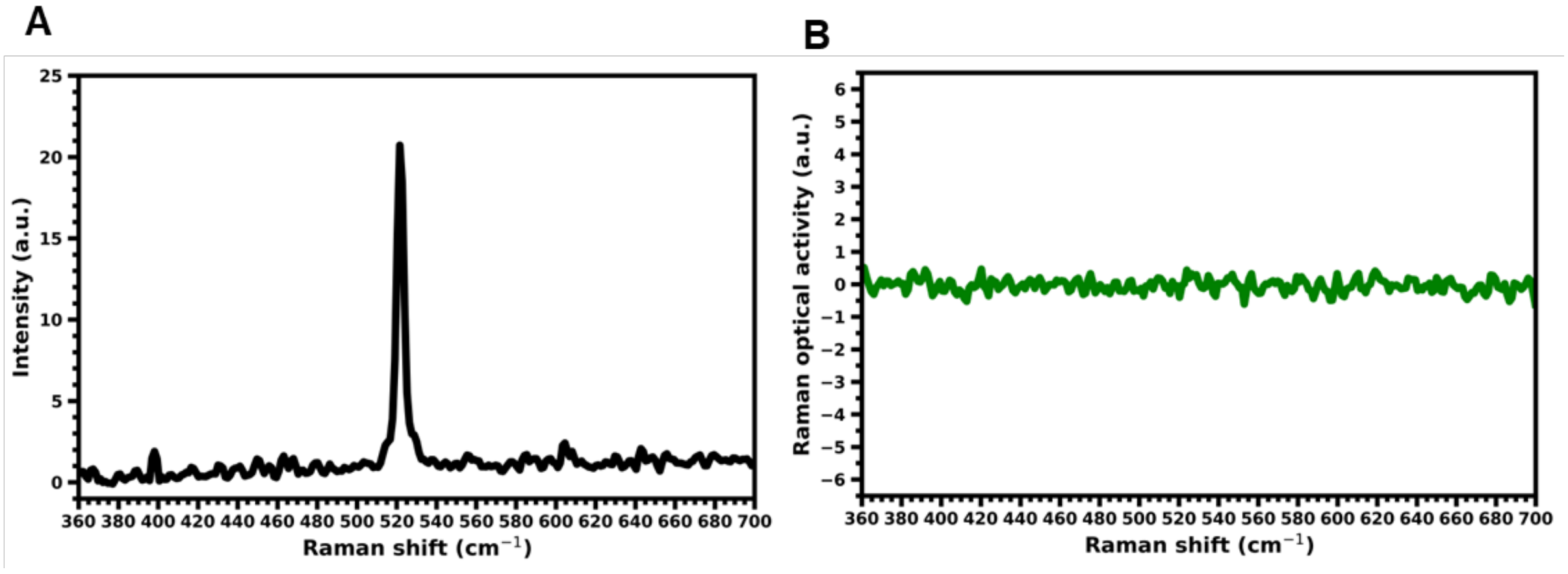

**Figure S4.** (A) Global Raman spectra and (B) global ROA spectra of the Au nanopattern.

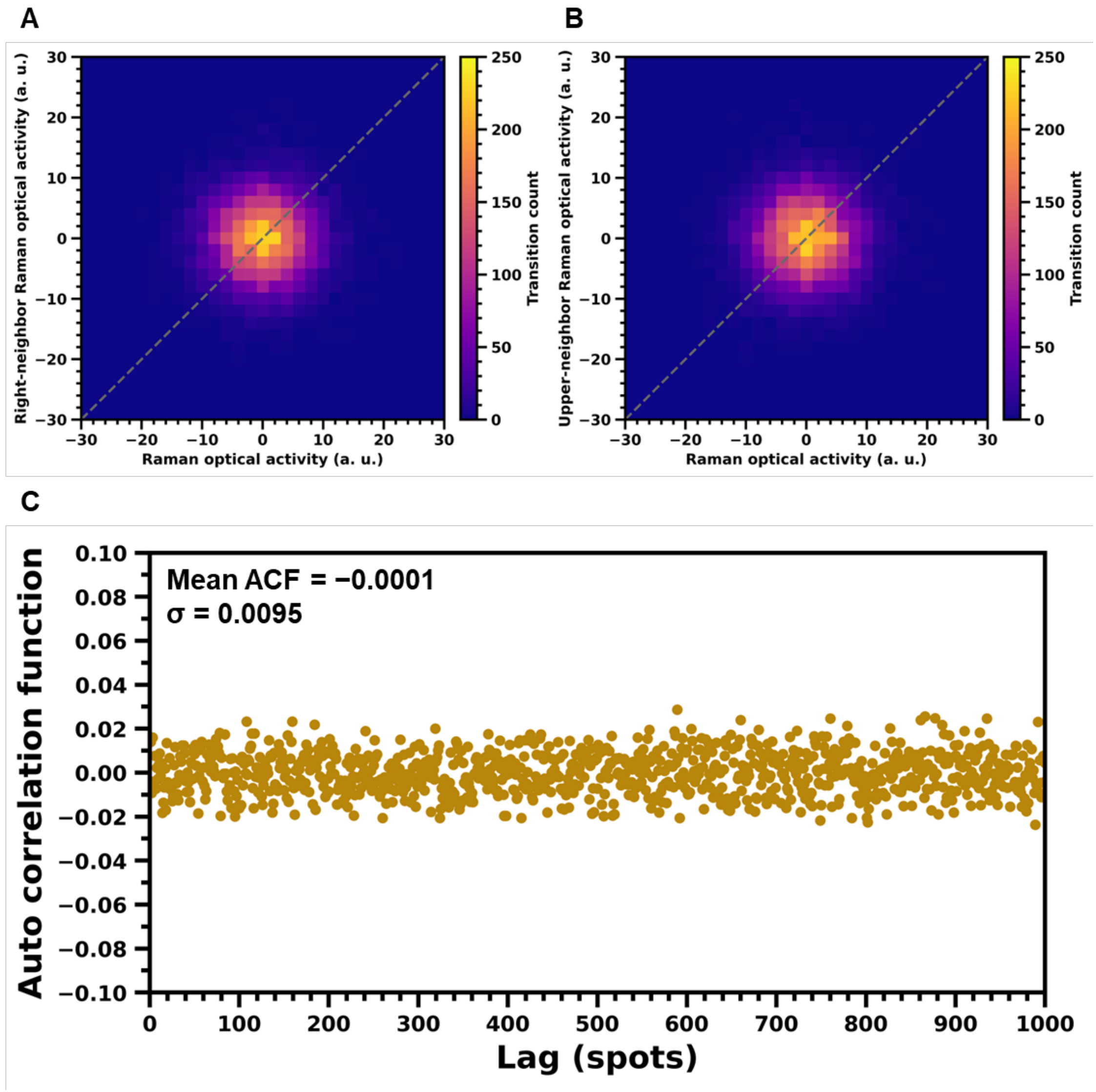

**Figure S5.** Statistical distribution and spatial independence of ROA signals in Au nanopatterns. Joint distribution maps showing the relationship between the ROA intensity of a reference spot and that of its (A) right-neighbor and (B) upper-neighbor spots. (C) Spatial auto correlation function (ACF) plotted as a function of spatial lag to evaluate the global independence of the spots.

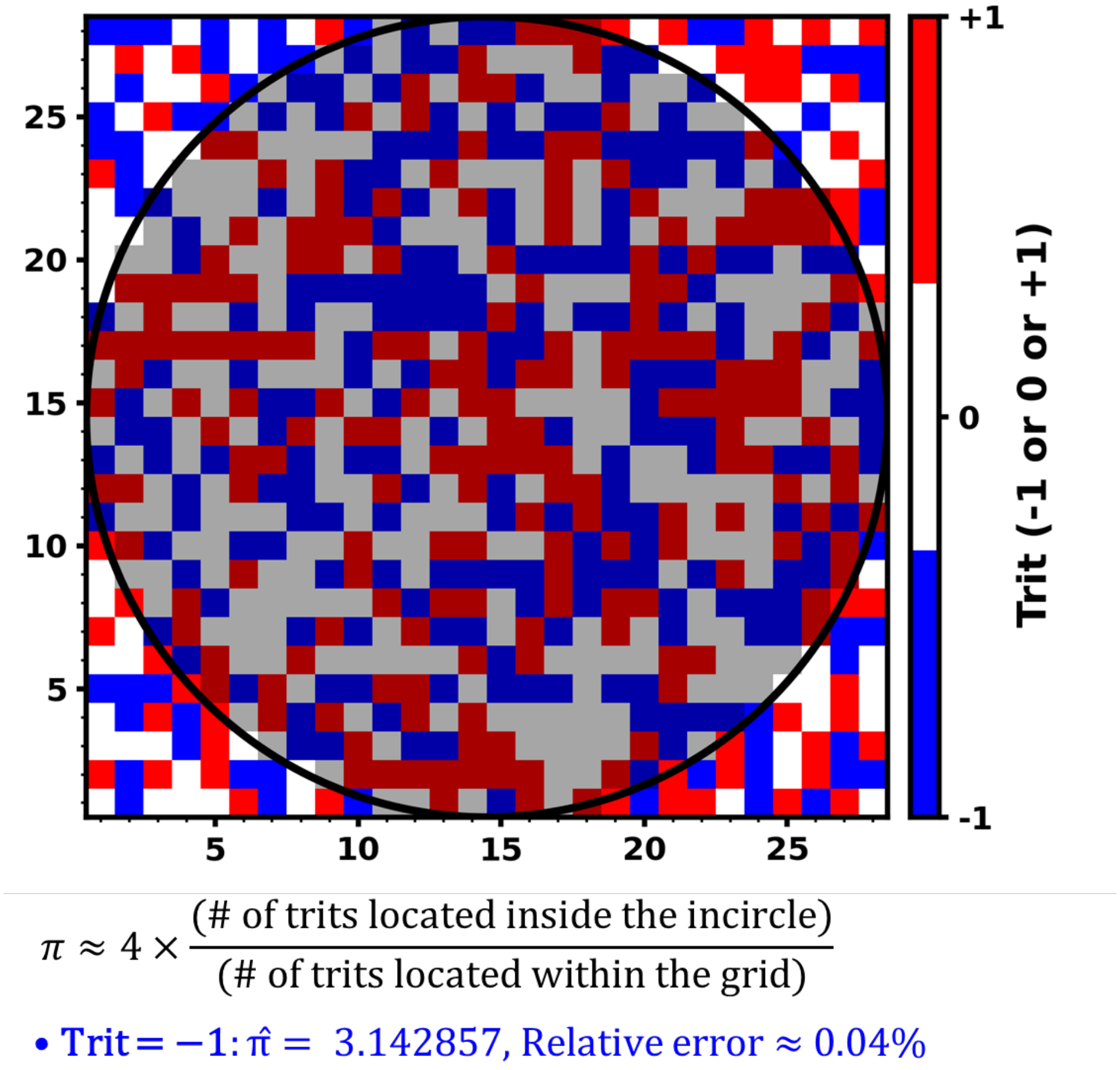

**Figure S6.** Intuitive verification of spatial randomness via Monte Carlo π estimation. The simulation was performed using the magnified 28 × 28 ternary grid displayed in Figure 3C. An inscribed circle was superimposed on the grid, and π was approximated based on the ratio of trits located within the circle for each distinct trit value.

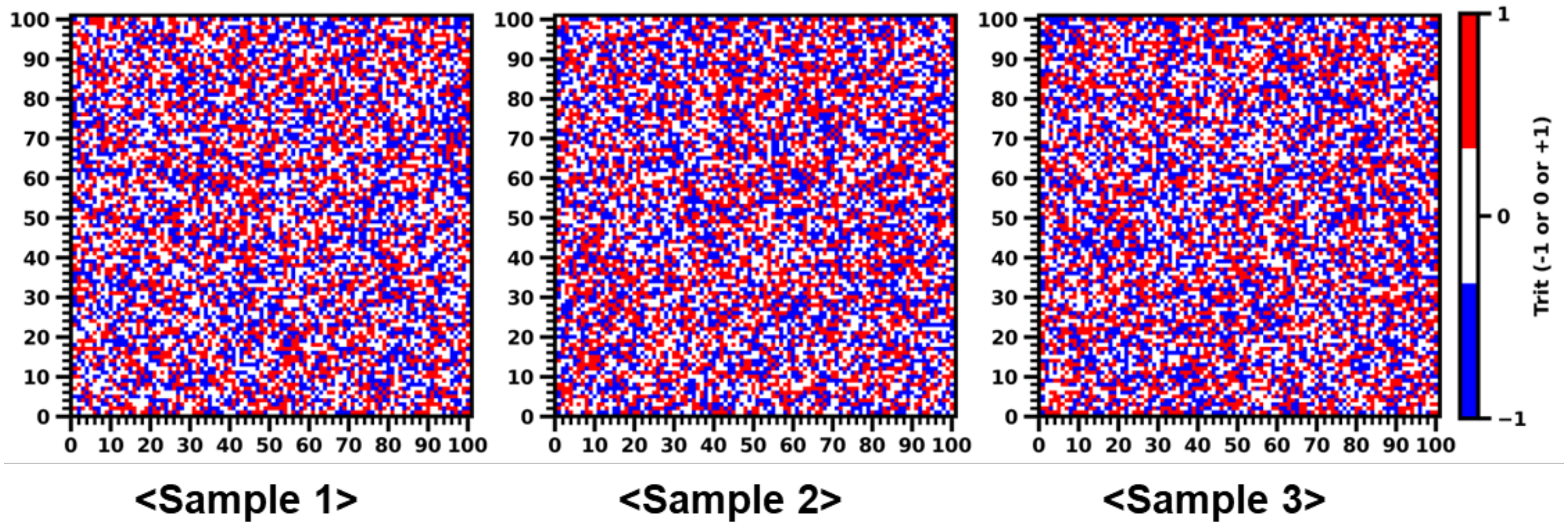

**Figure S7. Ternary heatmaps generated from three physically distinct Au nanopattern samples.** The resulting heatmaps displayed distinct, non-repeating patterns.

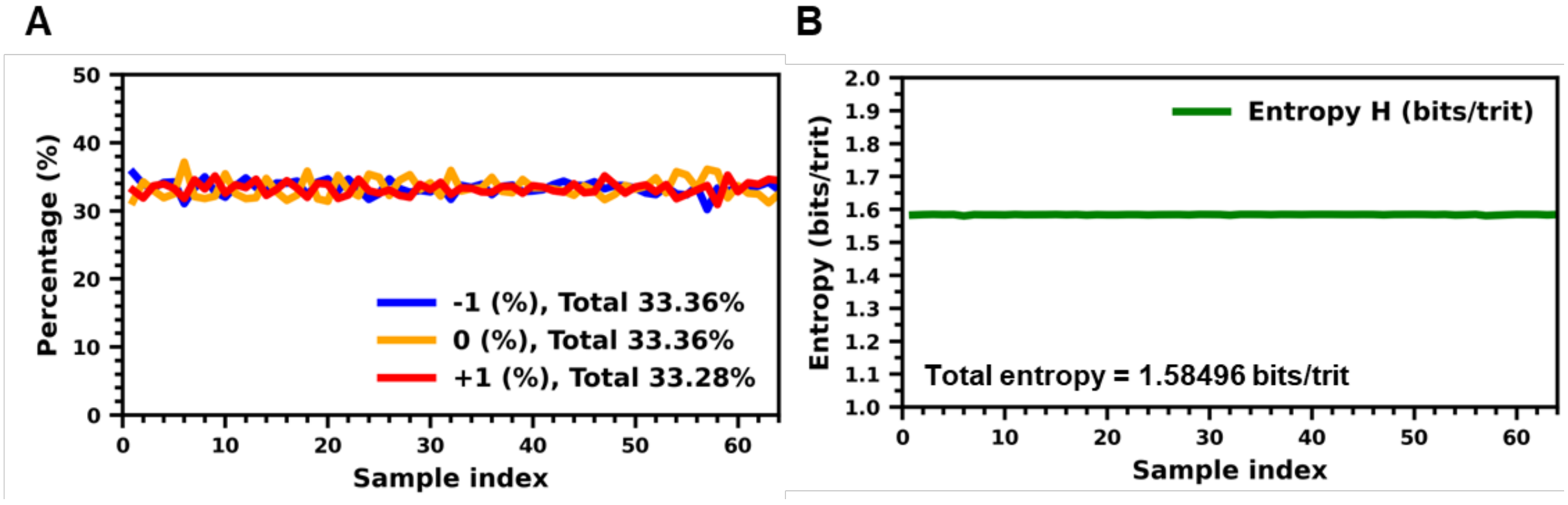

**Figure S8.** Unpredictable occurrence of each trit (−1/0/+1) value across 640,000 trits. (A) Sample-to-sample trit occupancy used to evaluate symbol balance across independently fabricated Au nanopattern samples. (B) Shannon entropy per sample used to quantify the information content of the ternary output.

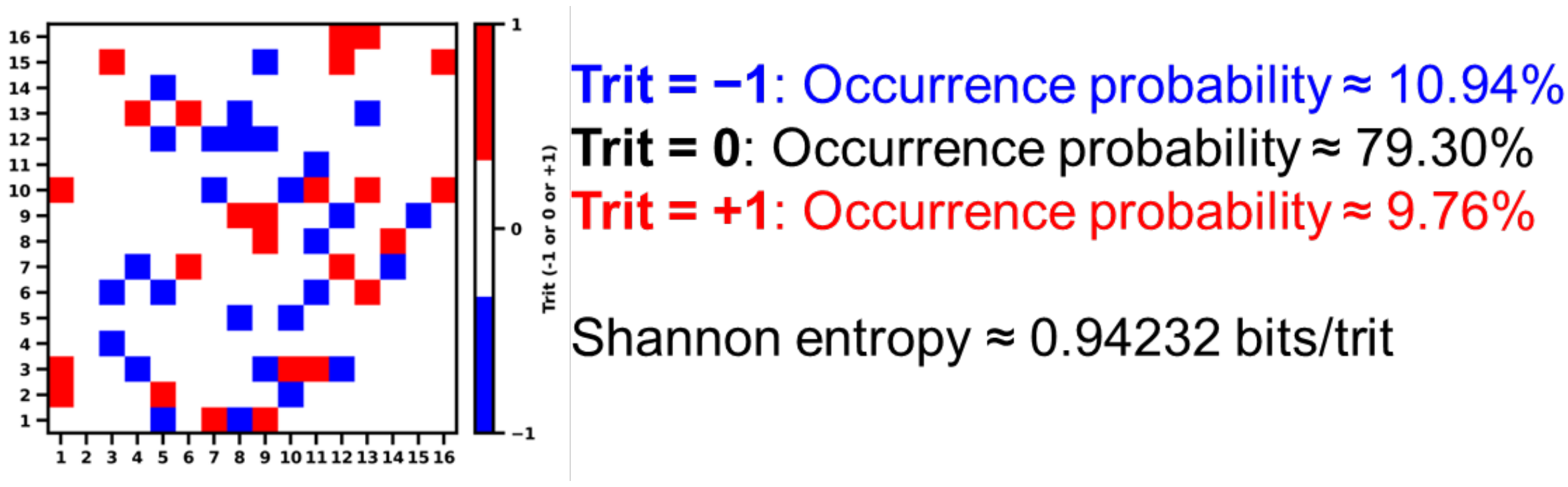

**Figure S9.** Control experiment using a bare Si wafer. A ternary map and statistical analysis were generated from a bare Si wafer without Au nanopatterns using the same extraction method as the main samples. Unlike the Au nanopattern samples, the bare wafer exhibits a distribution heavily biased toward '0'.

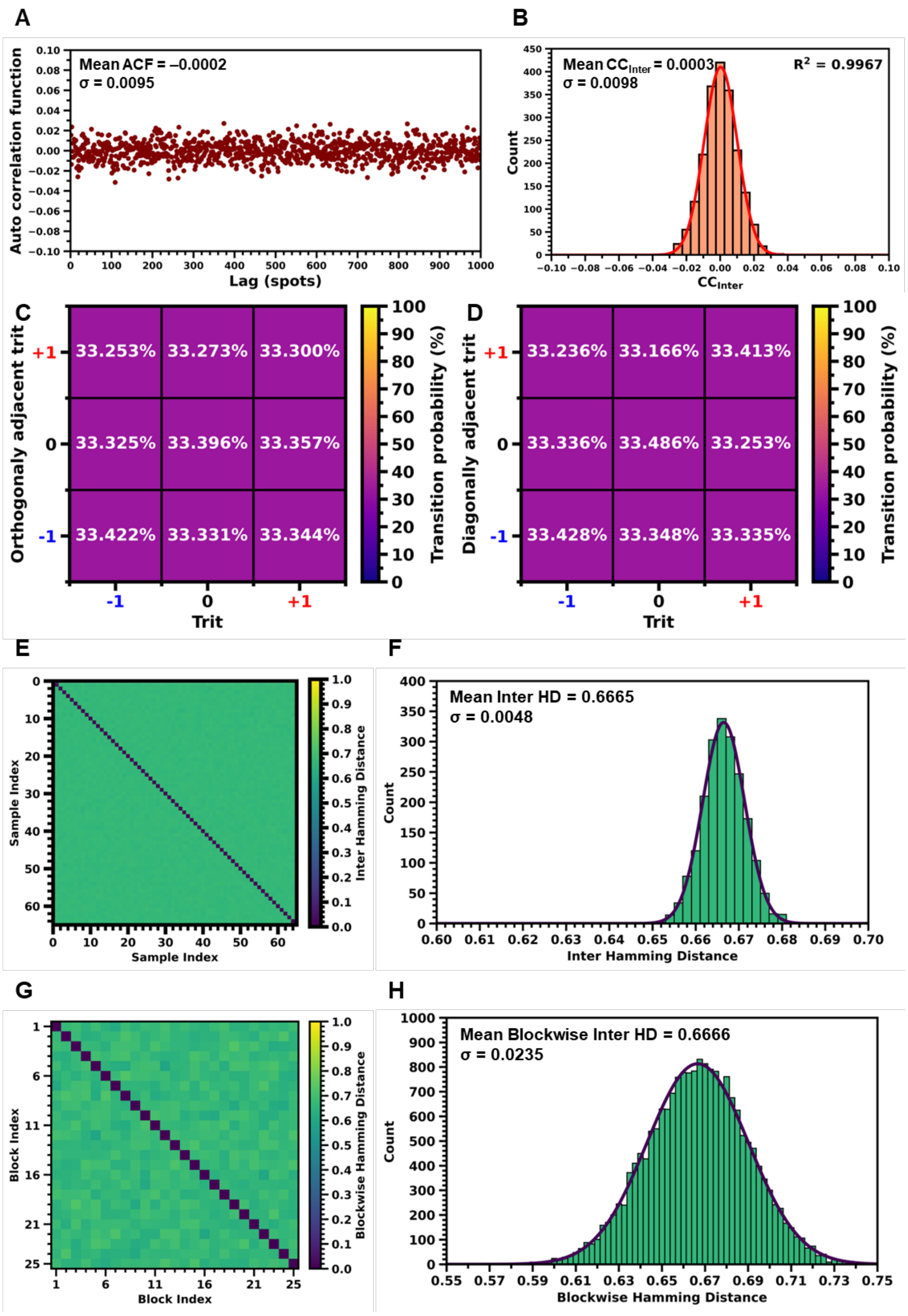

**Figure S10.** Statistical metrics to assess the quality of the chiroptical ternary entropy harvester. (A) Autocorrelation function (ACF) used to probe intra-sequence (spatial) correlation within the trit stream. (B) Distribution of inter-sample correlation coefficients used to assess linear dependence between sequences acquired from distinct samples. Heatmaps displaying the probability of finding a specific trit value adjacent to a center trit in (C) orthogonal and (D) diagonal directions. (E) Pairwise inter-sample normalized Hamming distance matrix used to quantify trit-level dissimilarity among sample-derived sequences. (F) Corresponding inter-sample Hamming-distance distribution. (G) Blockwise normalized Hamming distance matrix obtained by partitioning a single 100 × 100 trit map into 25 non-overlapping 20 × 20 blocks, used to evaluate trit-level independence across spatially separated regions within one map. (H) Corresponding blockwise Hamming-distance distribution.

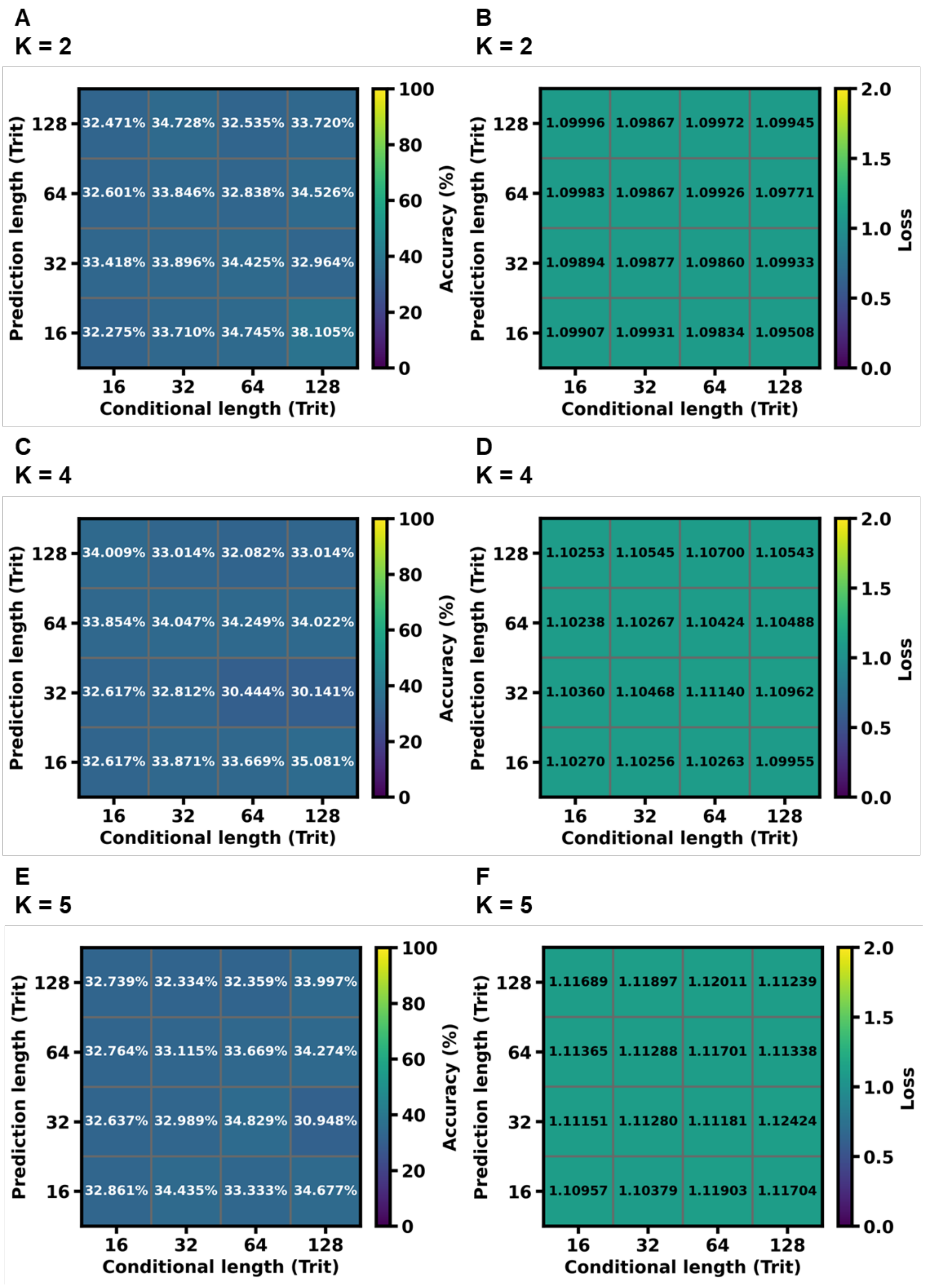

**Figure S11.** Markov-order dependence of predictability analysis for the chiroptical ternary entropy harvester output. Markov predictor results for K = 2: (A) validation accuracy and (B) per-trit categorical cross-entropy (loss) shown as a function of conditional length and prediction length. Markov predictor results for K = 4: (C) validation accuracy and (D) loss under the same conditions. Markov predictor results for K = 5: (E) validation accuracy and (F) loss under the same conditions.

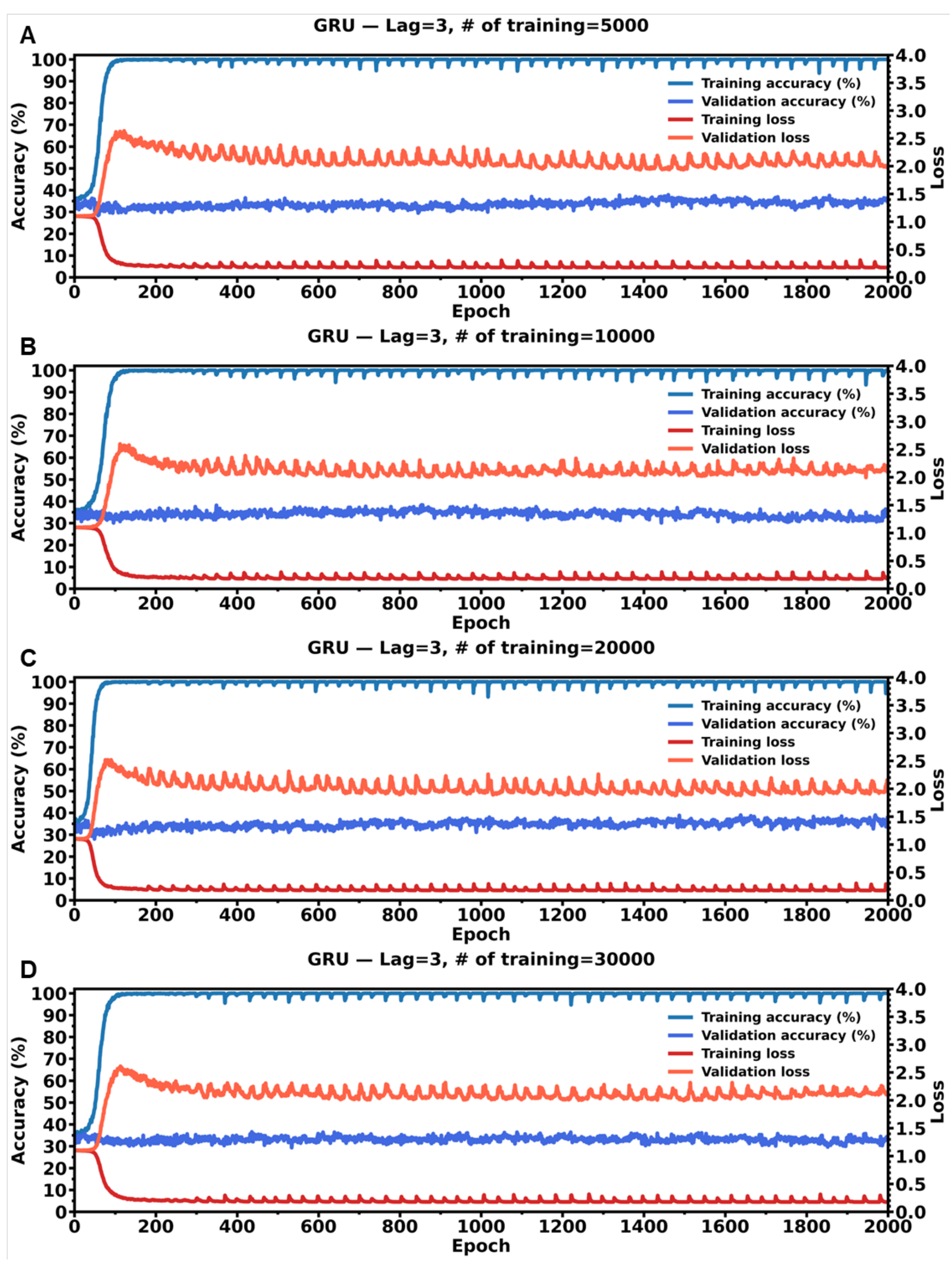

**Figure S12.** GRU training and validation learning curves at lag = 3. Training/validation accuracy and per-trit categorical cross-entropy are plotted as a function of epoch for different training-set sizes: (A) 5,000, (B) 10,000, (C) 20,000, and (D) 30,000 sequences.

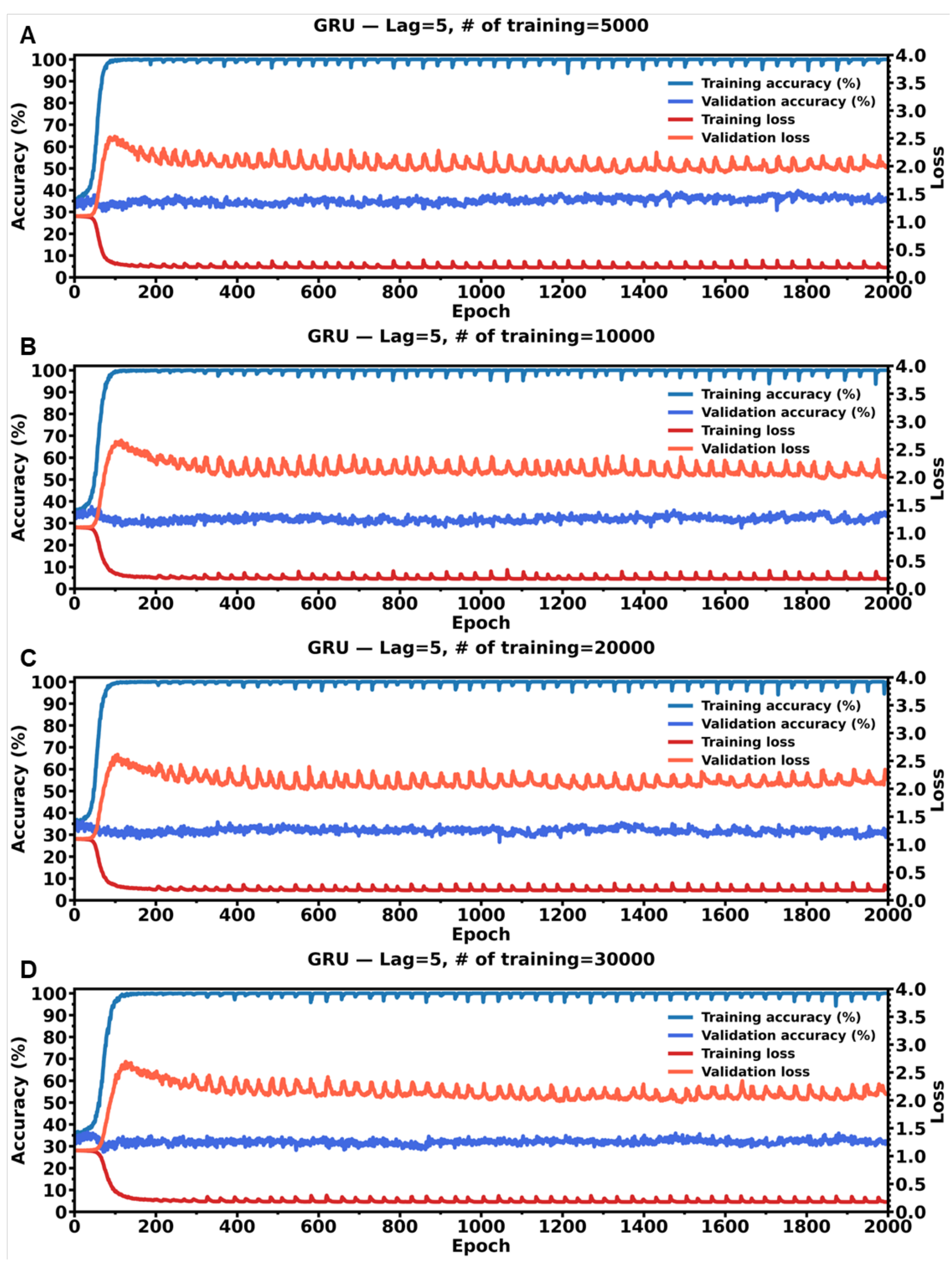

**Figure S13.** GRU training and validation learning curves at lag = 5. Training/validation accuracy and per-trit categorical cross-entropy are plotted as a function of epoch for different training-set sizes: (A) 5,000, (B) 10,000, (C) 20,000, and (D) 30,000 sequences.

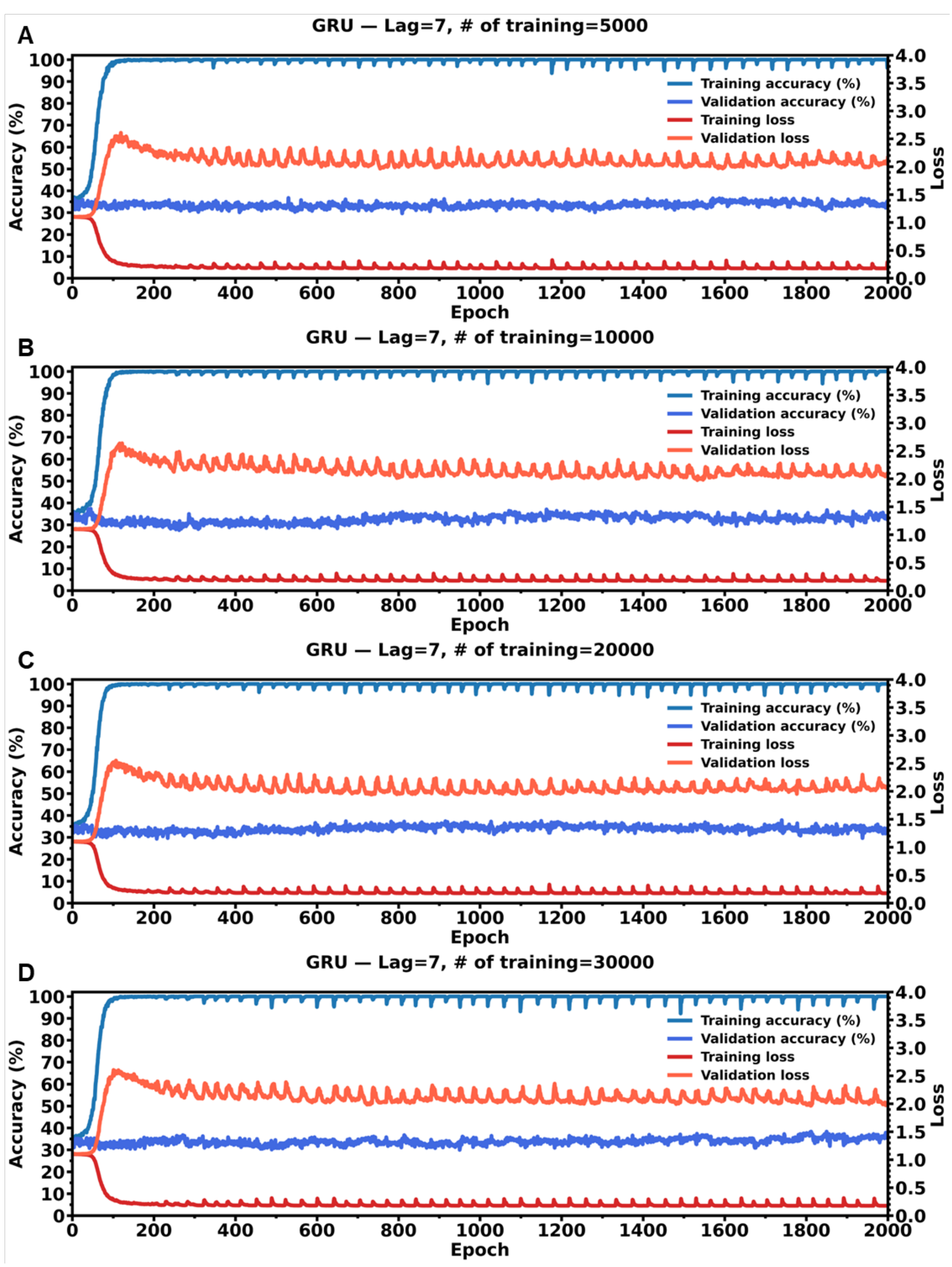

**Figure S14.** GRU training and validation learning curves at lag = 7. Training/validation accuracy and per-trit categorical cross-entropy are plotted as a function of epoch for different training-set sizes: (A) 5,000, (B) 10,000, (C) 20,000, and (D) 30,000 sequences.

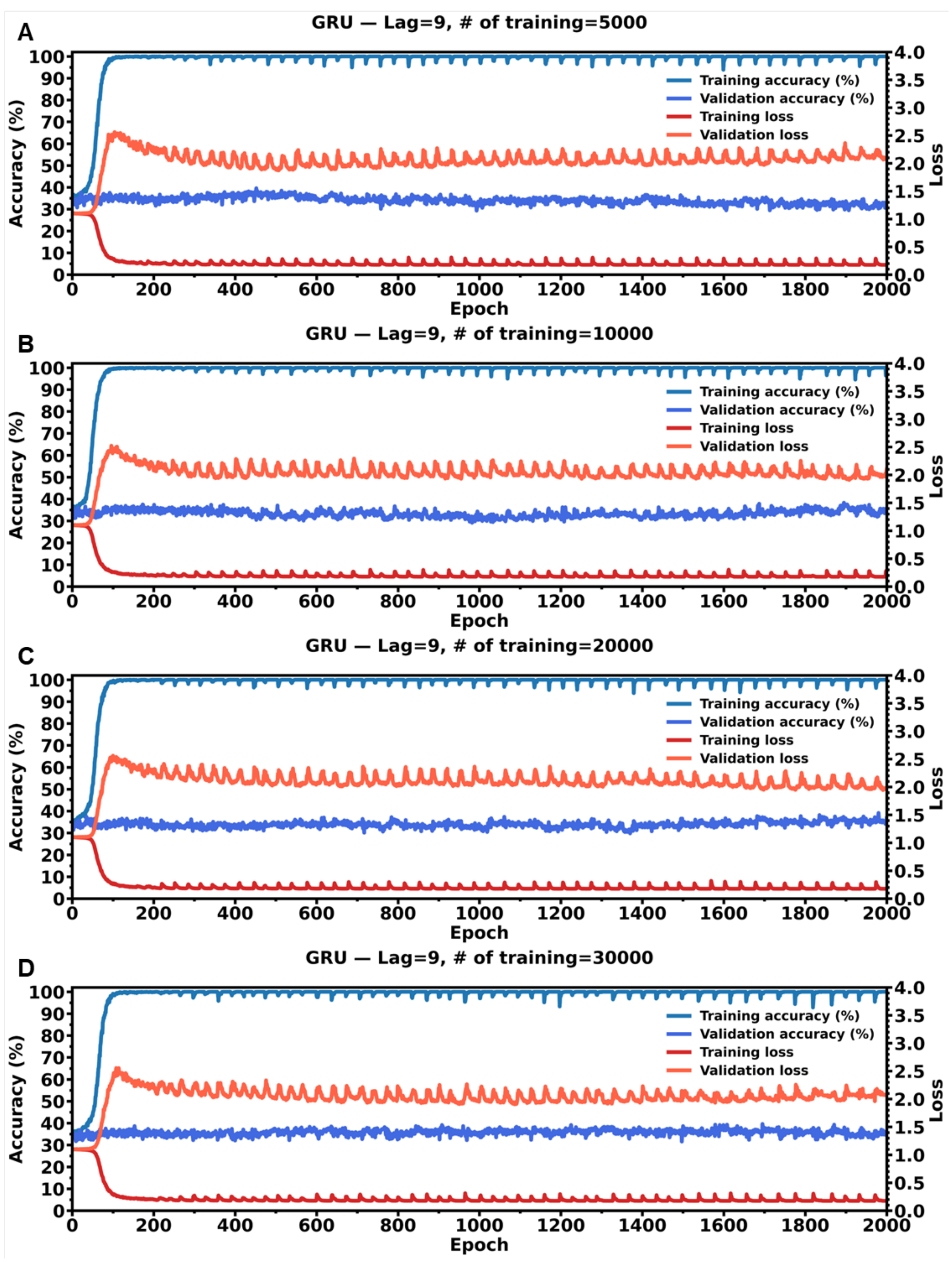

**Figure S15.** GRU training and validation learning curves at lag = 9. Training/validation accuracy and per-trit categorical cross-entropy are plotted as a function of epoch for different training-set sizes: (A) 5,000, (B) 10,000, (C) 20,000, and (D) 30,000 sequences.

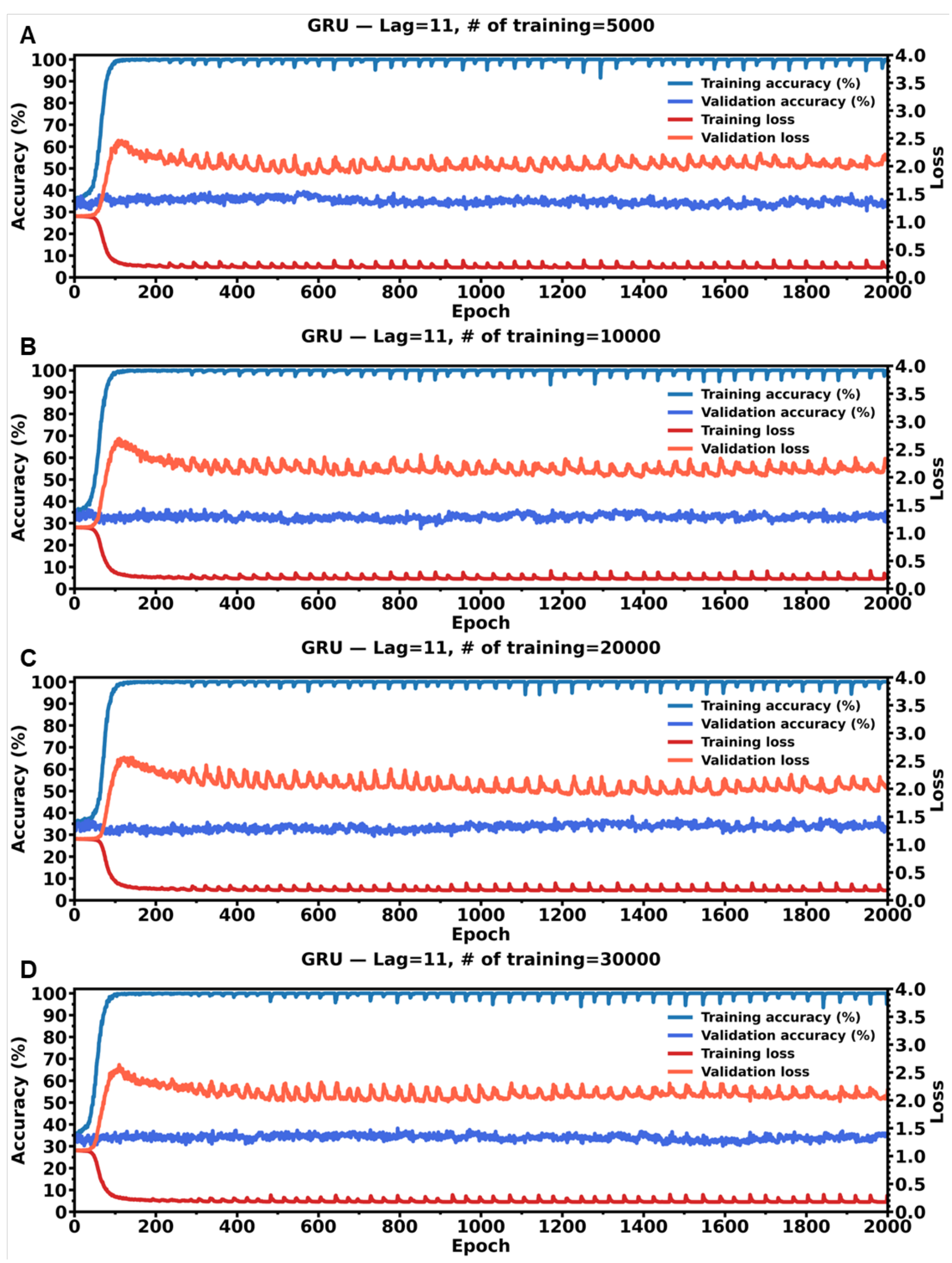

**Figure S16.** GRU training and validation learning curves at lag = 11. Training/validation accuracy and per-trit categorical cross-entropy are plotted as a function of epoch for different training-set sizes: (A) 5,000, (B) 10,000, (C) 20,000, and (D) 30,000 sequences.

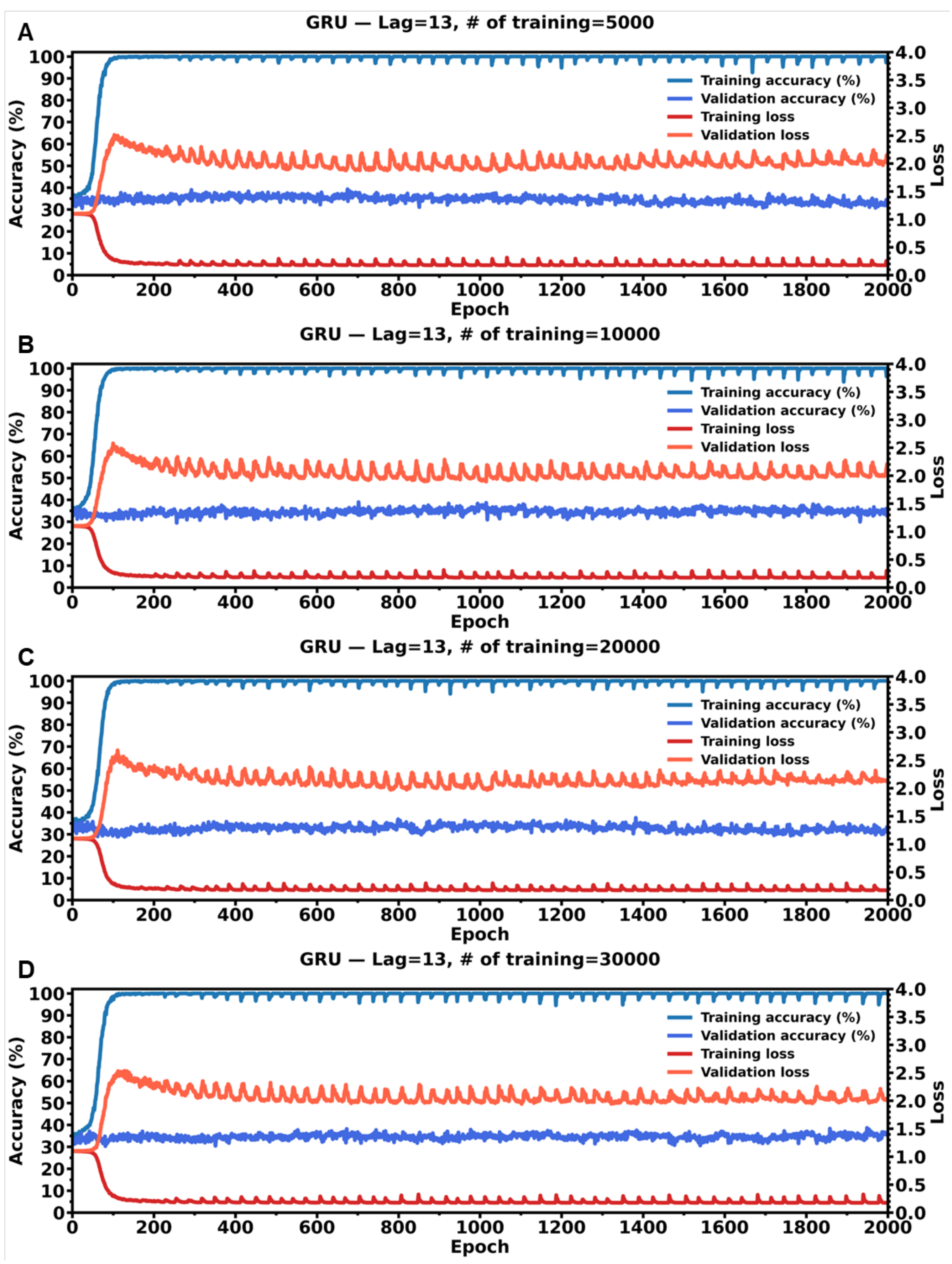

**Figure S17.** GRU training and validation learning curves at lag = 13. Training/validation accuracy and per-trit categorical cross-entropy are plotted as a function of epoch for different training-set sizes: (A) 5,000, (B) 10,000, (C) 20,000, and (D) 30,000 sequences.

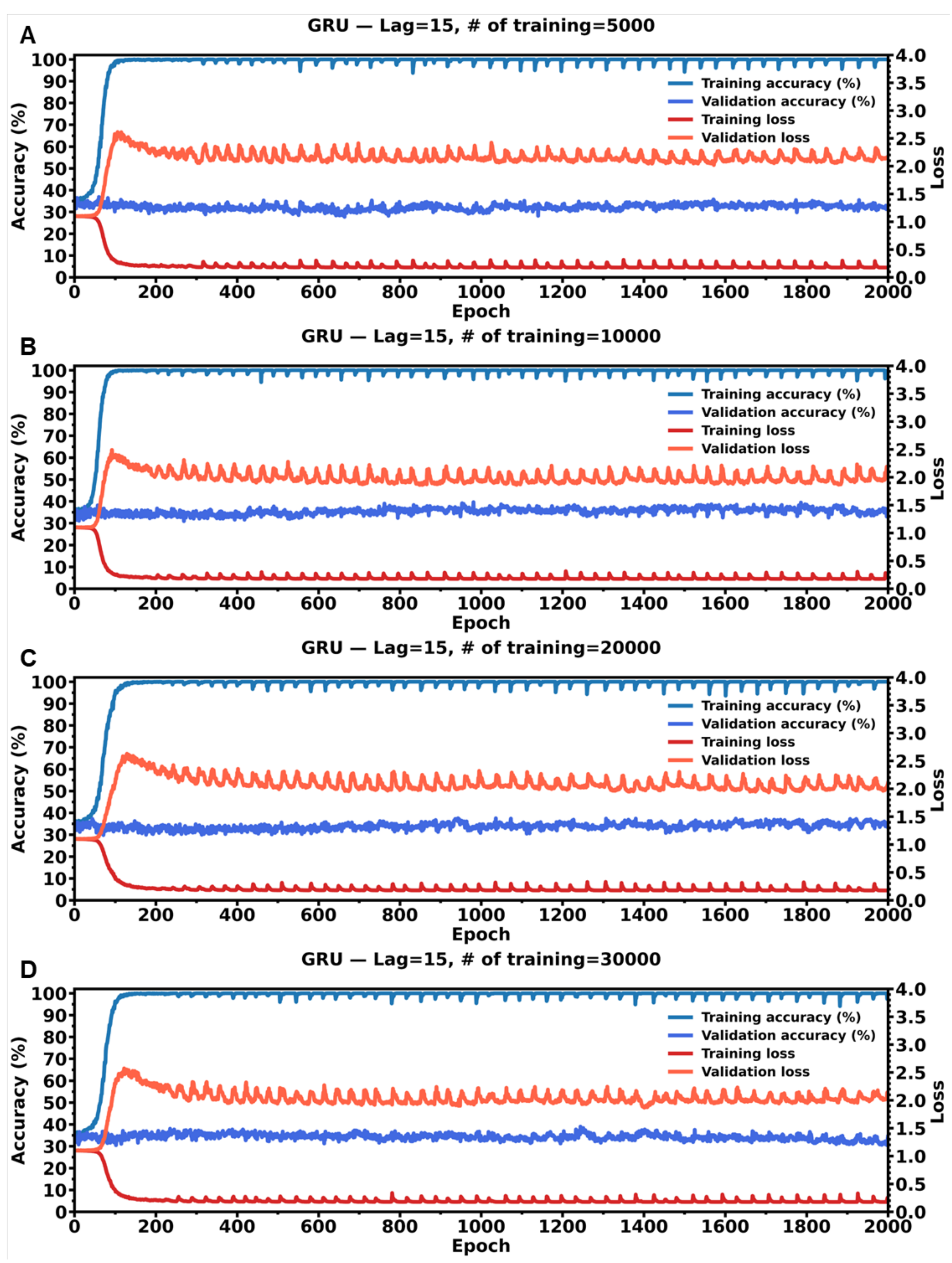

**Figure S18.** GRU training and validation learning curves at lag = 15. Training/validation accuracy and per-trit categorical cross-entropy are plotted as a function of epoch for different training-set sizes: (A) 5,000, (B) 10,000, (C) 20,000, and (D) 30,000 sequences.

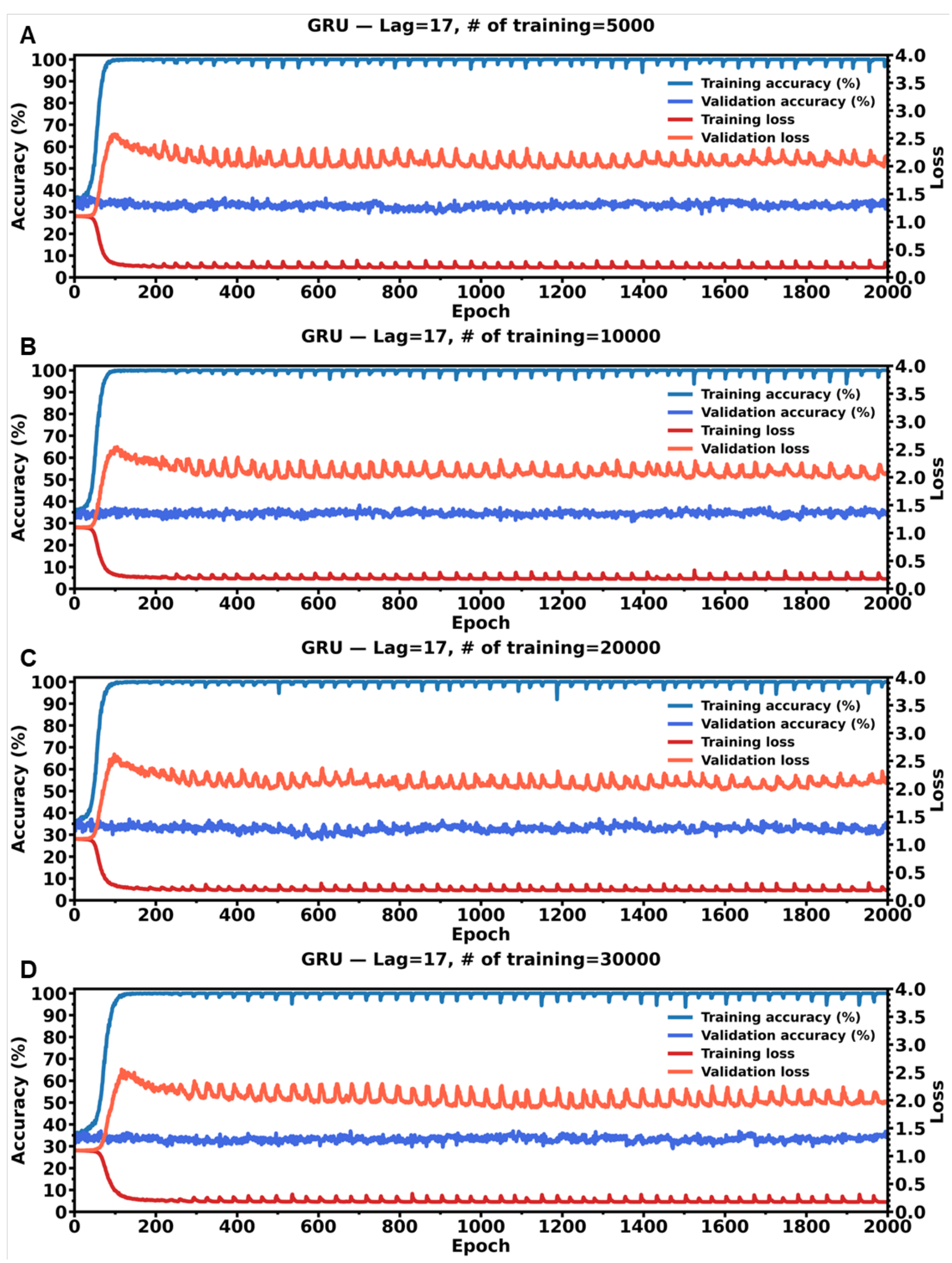

**Figure S19.** GRU training and validation learning curves at lag = 17. Training/validation accuracy and per-trit categorical cross-entropy are plotted as a function of epoch for different training-set sizes: (A) 5,000, (B) 10,000, (C) 20,000, and (D) 30,000 sequences.

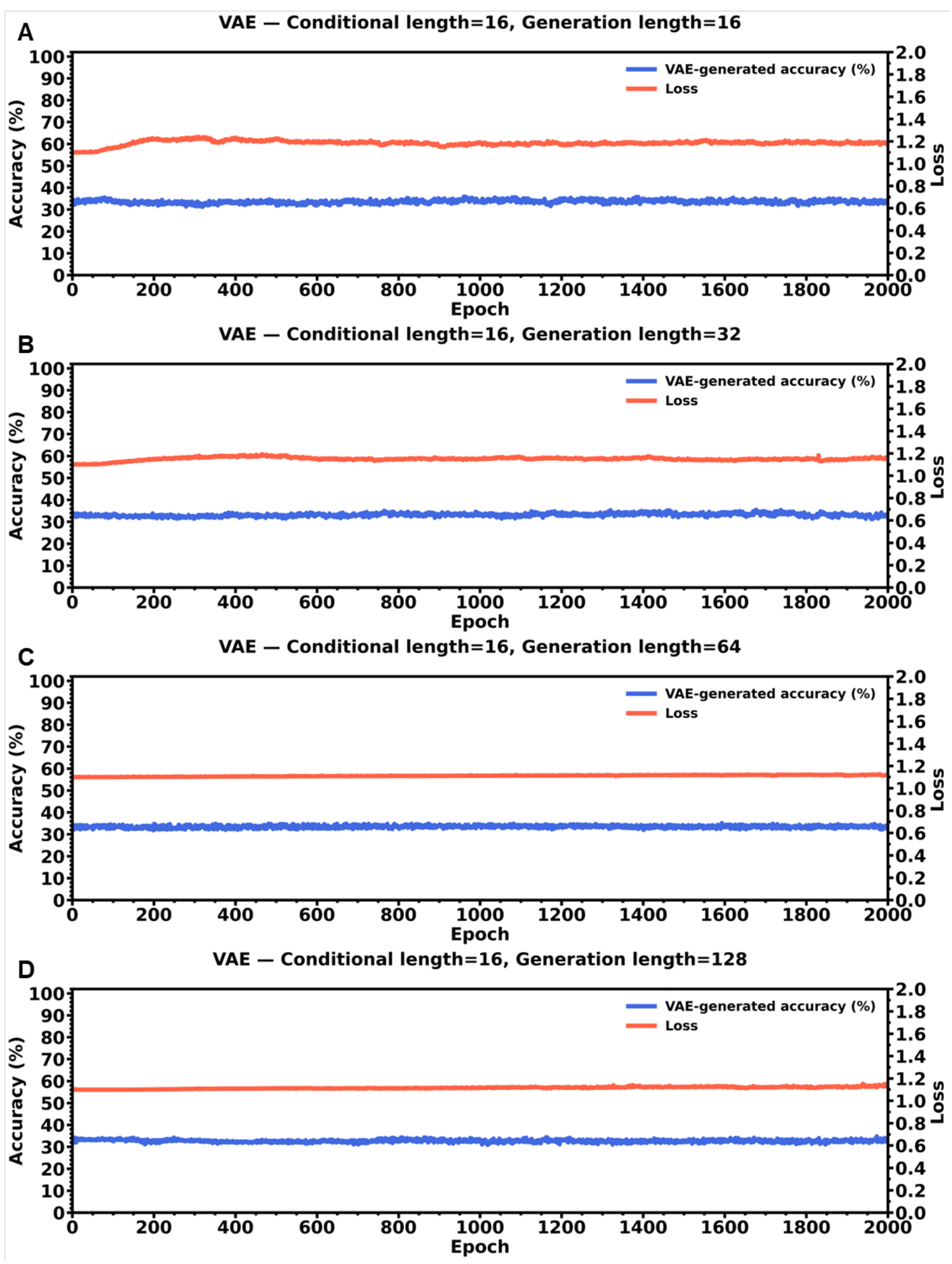

**Figure S20.** VAE learning curves for conditional generation at conditional length = 16. Validation accuracy and per-trit categorical cross-entropy are plotted as a function of epoch for different generation lengths: (A) 16, (B) 32, (C) 64, and (D) 128 trits.

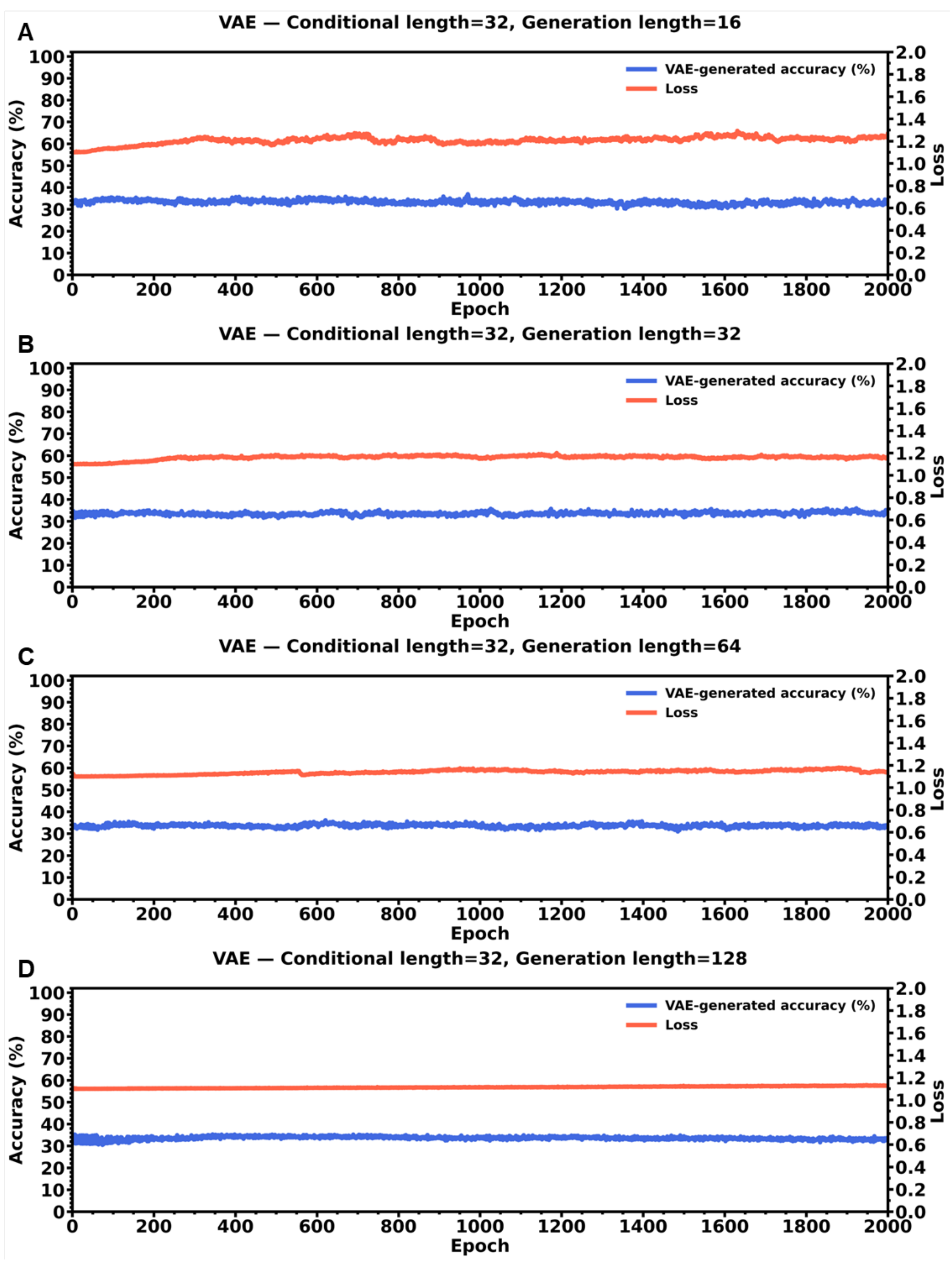

**Figure S21.** VAE learning curves for conditional generation at conditional length = 32. Validation accuracy and per-trit categorical cross-entropy are plotted as a function of epoch for different generation lengths: (A) 16, (B) 32, (C) 64, and (D) 128 trits.

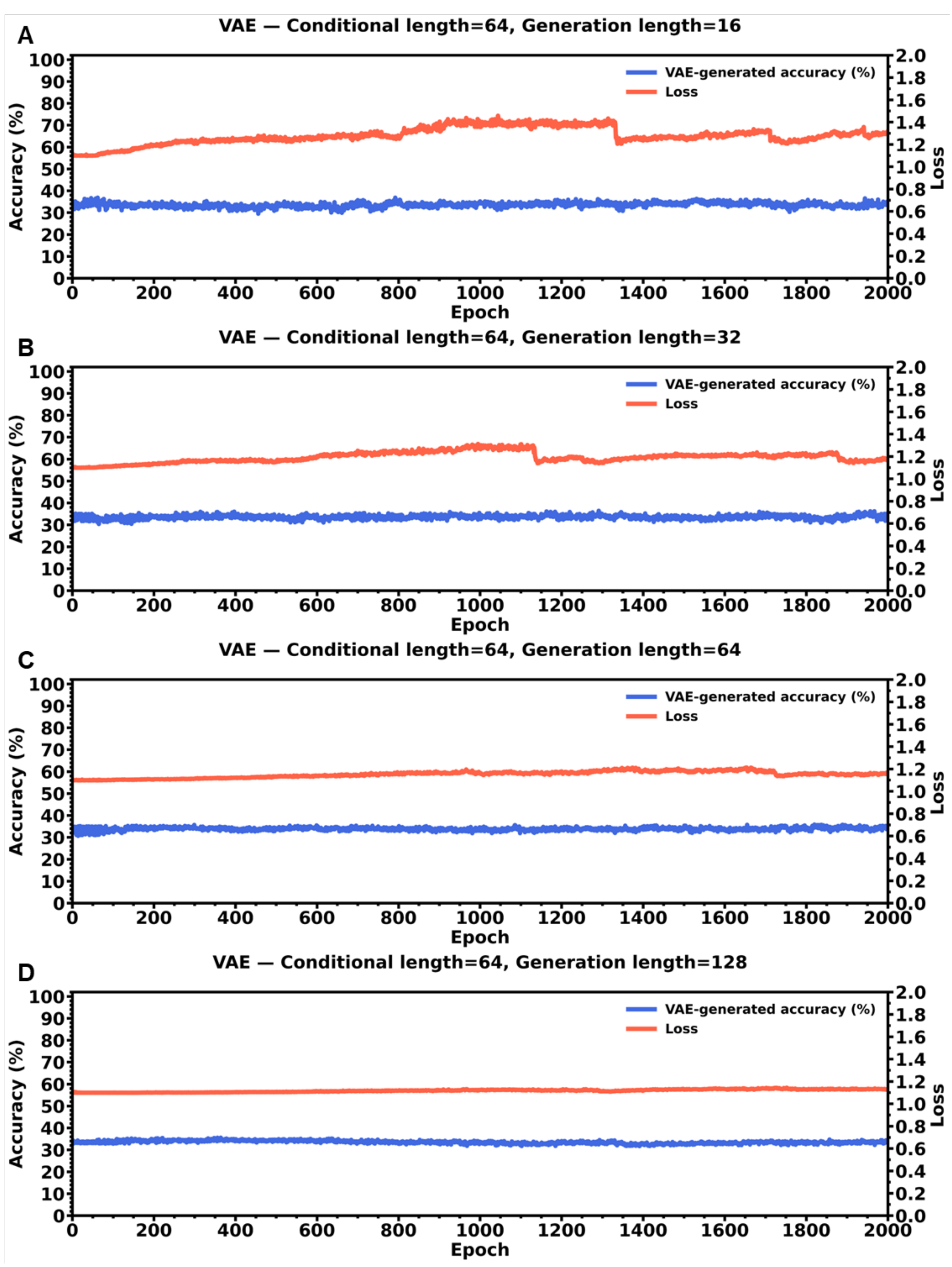

**Figure S22.** VAE learning curves for conditional generation at conditional length = 64. Validation accuracy and per-trit categorical cross-entropy are plotted as a function of epoch for different generation lengths: (A) 16, (B) 32, (C) 64, and (D) 128 trits.

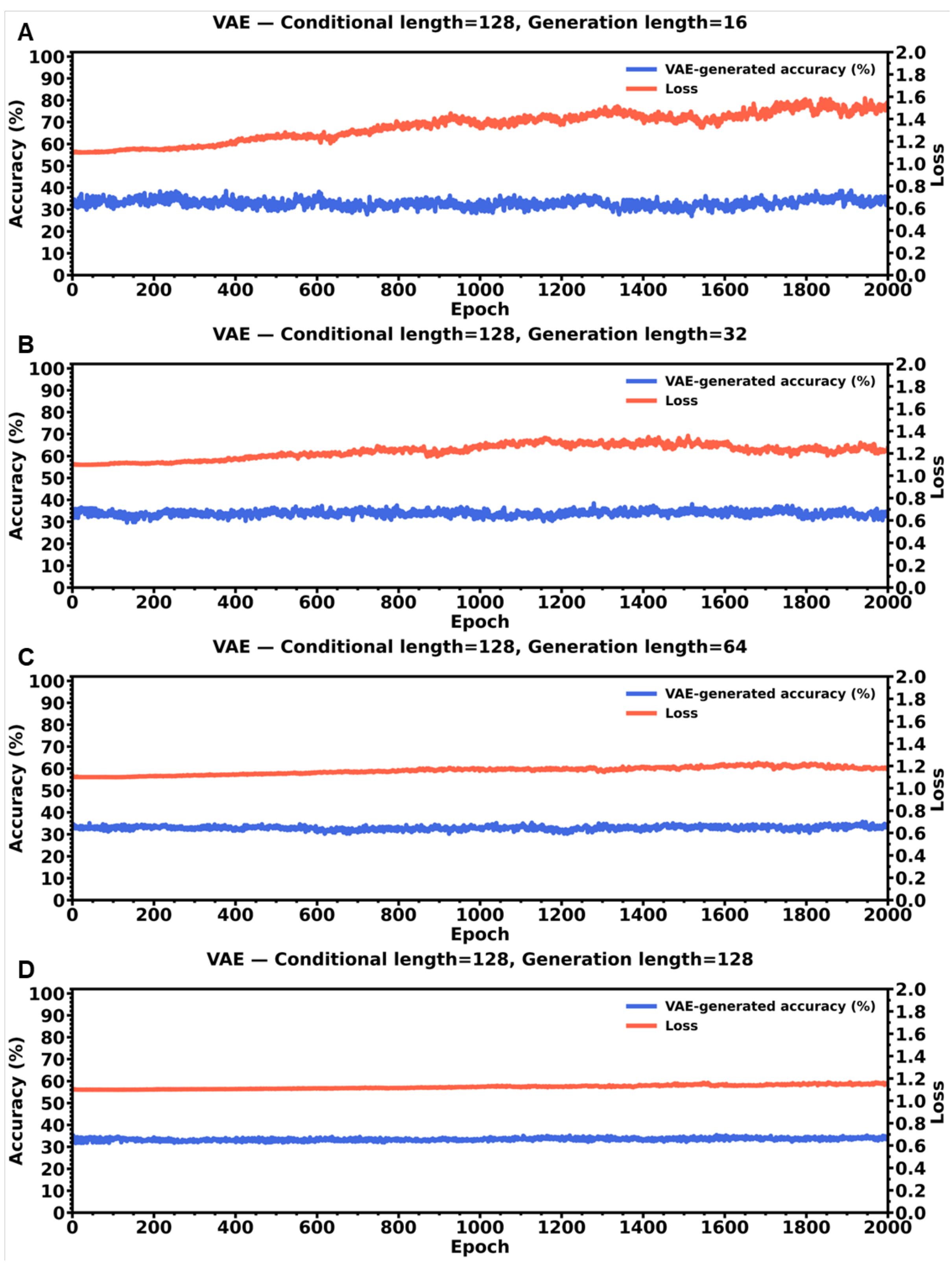

**Figure S23.** VAE learning curves for conditional generation at conditional length = 128. Validation accuracy and per-trit categorical cross-entropy are plotted as a function of epoch for different generation lengths: (A) 16, (B) 32, (C) 64, and (D) 128 trits.

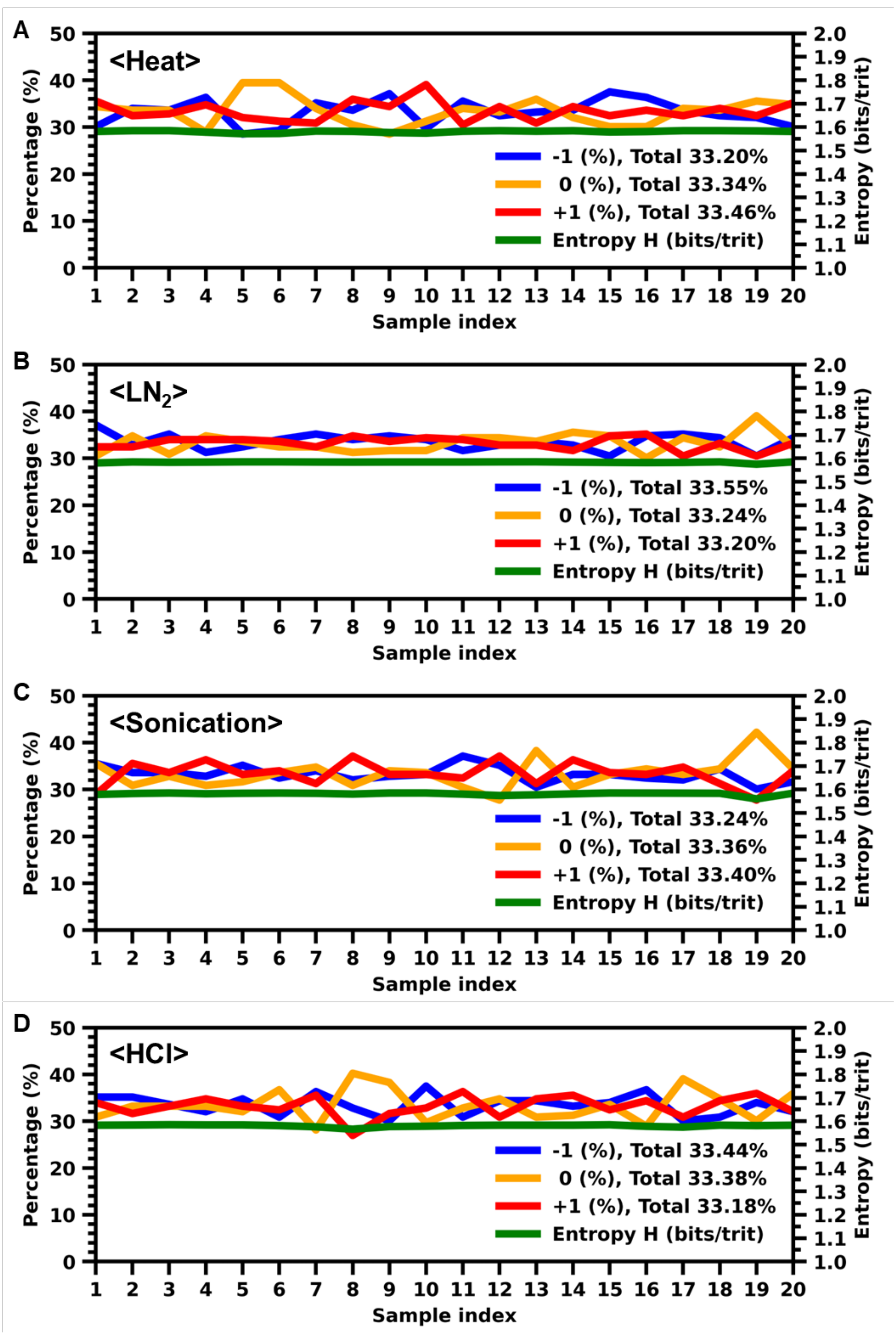

**Figure S24.** Statistical analysis for robustness assessment of the Au nanopattern-on-Si platform for ternary entropy harvester operation under harsh environments. Trit occupancy and Shannon entropy per sample after (A) heating at 150 °C, (B) cryogenic exposure in liquid nitrogen (−196 °C), (C) ultrasonication in water, and (D) immersion in 1 M HCl for 1 h.

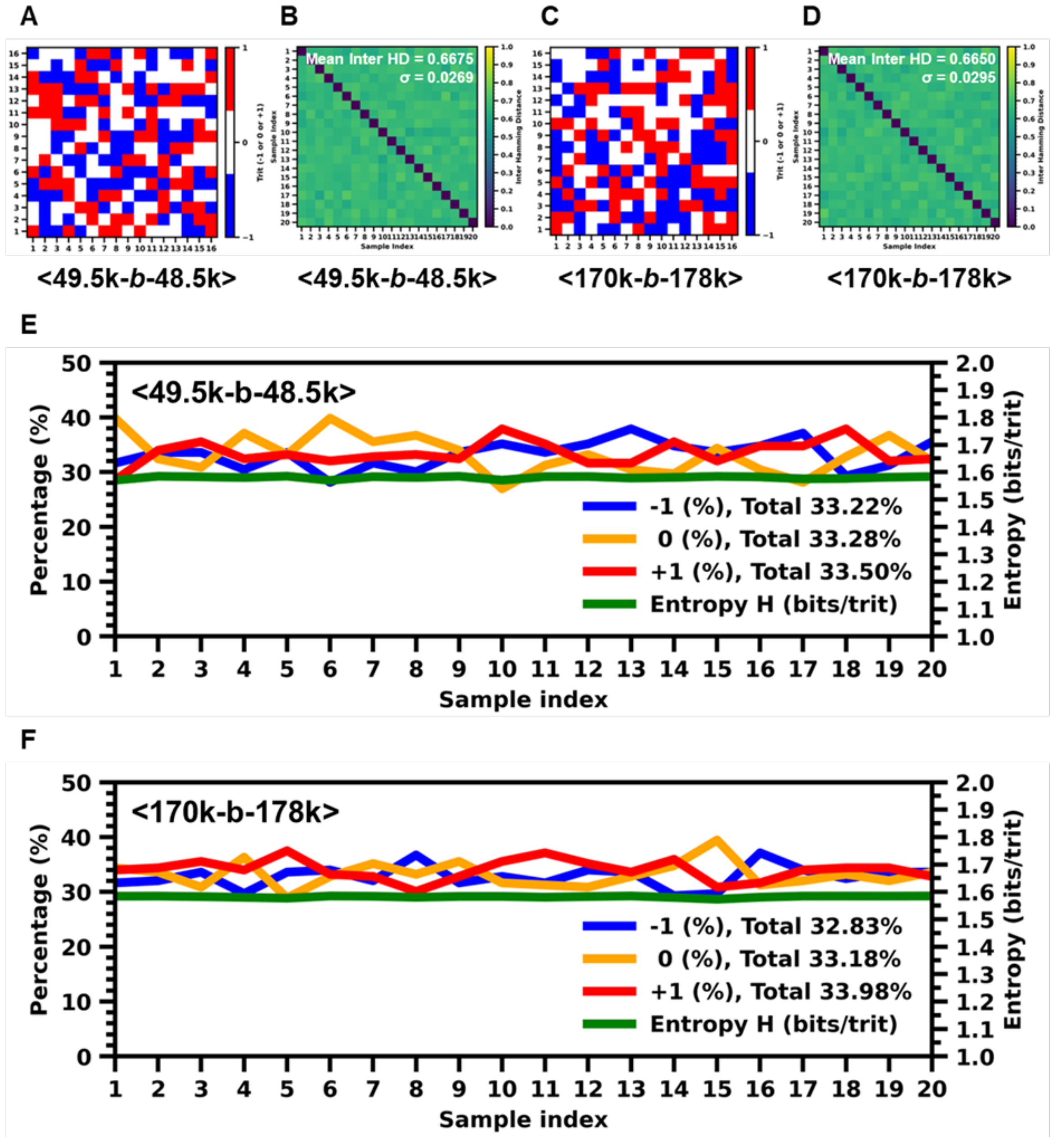

**Figure S25.** Statistical metrics to assess the quality of the chiroptical ternary entropy harvester using Au nanopatterns fabricated by PS-*b*-PMMA with various molecular weight. (A) Ternary heatmap and (B) inter-sample normalized Hamming distance matrix from Au nanopattern samples fabricated by PS-*b*-PMMA with the molecular weight of 49.5k-*b*-48.5k. (C) Ternary heatmap and (D) inter-sample normalized Hamming distance matrix from Au nanopattern samples fabricated by PS-*b*-PMMA with the molecular weight of 170k-*b*-178k. Trit occupancies and the corresponding Shannon entropy extracted from Au nanopatterns fabricated by (E) PS-*b*-PMMA with the molecular weight of 49.5k-*b*-48.5k and (F) PS-*b*-PMMA with the molecular weight of 170k-*b*-178k.

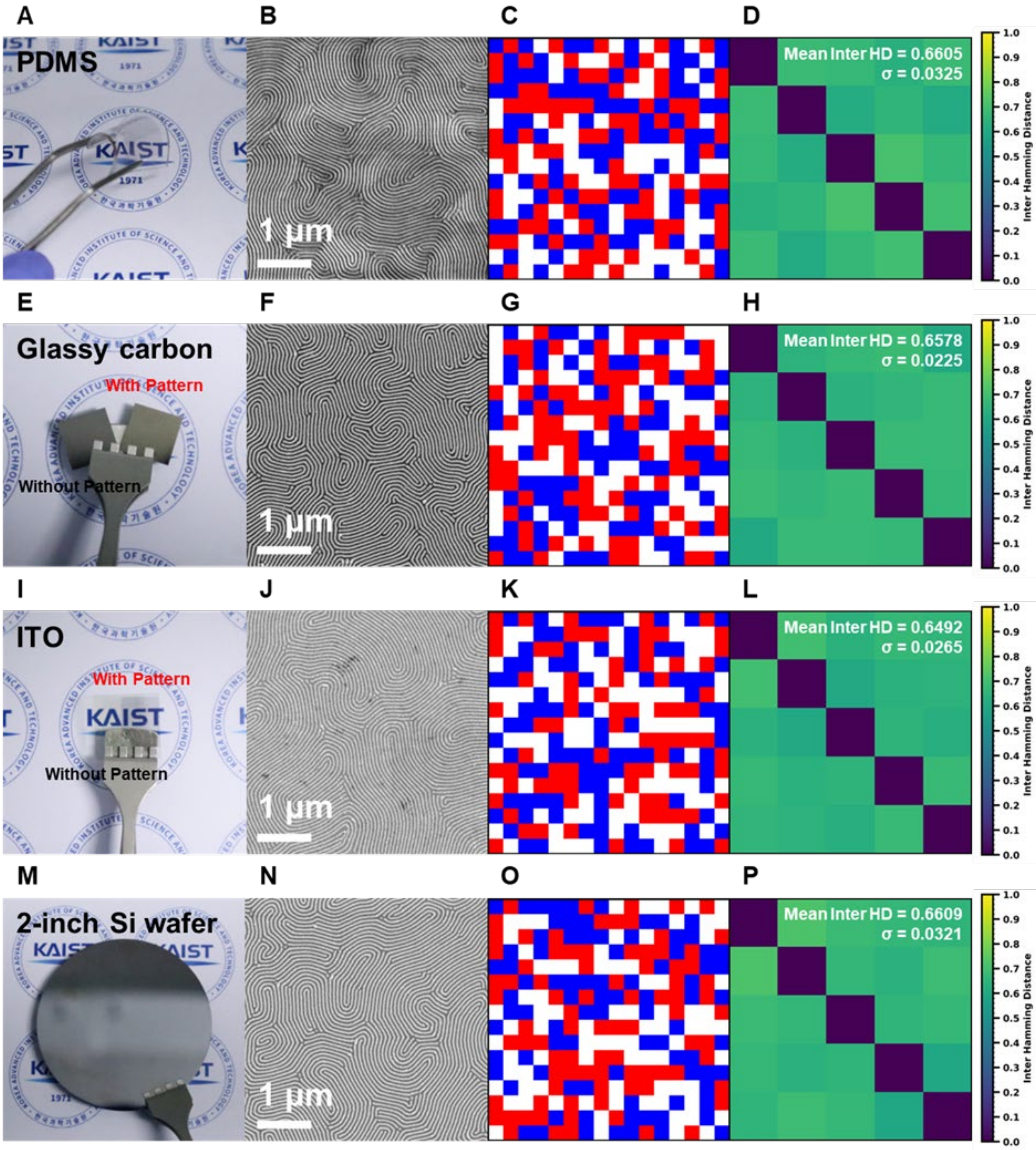

**Figure S26.** Substrate versatility and wafer-scale scalability of the chiroptical ternary entropy harvesting system. Au nanopatterns fabricated on a PDMS substrate: (A) photograph, (B) SEM image, (C) extracted ternary heatmap, and (D) inter-sample normalized Hamming-distance matrix. Au nanopatterns fabricated on a glassy carbon substrate: (E) photograph, (F) SEM image, (G) extracted ternary heatmap, and (H) inter-sample normalized Hamming-distance matrix. Au nanopatterns fabricated on an ITO substrate: (I) photograph, (J) SEM image, (K) extracted ternary heatmap, and (L) inter-sample normalized Hamming-distance matrix. Wafer-scale demonstration on a 2-inch Si wafer: (M) photograph, (N) SEM image, (O) extracted ternary heatmap, and (P) inter-sample normalized Hamming-distance matrix.

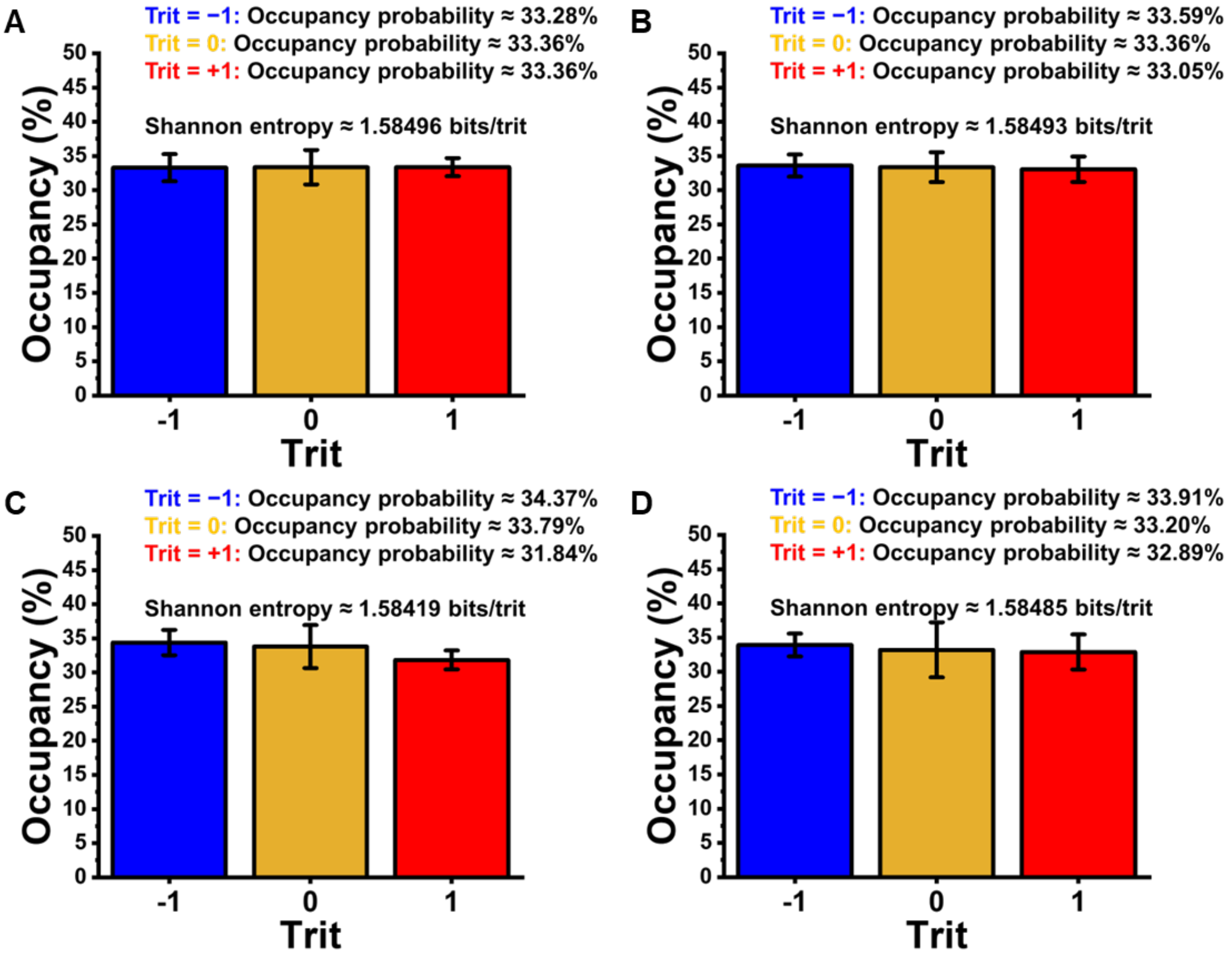

**Figure S27.** Trit occupancies and the corresponding Shannon entropy extracted from Au nanopatterns fabricated on (A) polydimethylsiloxane (PDMS), (B) glassy carbon, (C) indium tin oxide (ITO), and (D) a 2-inch Si wafer. Error bars represent the standard deviation across samples (n = 5 for each substrate).

**Supplementary Notes**

**Calculation of Raman Emission Intensity in Plasmonic Nanostructures**

Raman emission is strongly governed by the local electric near field at the optical focus.[1,2] In plasmonic nanostructures, the field-enhanced Raman emission measured at a spatial coordinate $\mathbf{r}$ can be expressed as follows.

$$I_{\mathrm{En}}(\mathbf{r}) = G_{\mathrm{En}}(\mathbf{r}) \times I_{\mathrm{ref}} \tag{S1}$$

, where $I_{\mathrm{En}}(\mathbf{r})$ is the measured Raman intensity under near-field enhancement at $\mathbf{r}$, $I_{\mathrm{ref}}$ is the Raman intensity in a reference environment without near-field enhancement, and $G_{\mathrm{En}}(\mathbf{r})$ is the Raman enhancement factor arising from near-field amplification at $\mathbf{r}$.[3] In this context, $G_{\mathrm{En}}(\mathbf{r})$ is a function of the local electric near field at $\mathbf{r}$ and can be expressed as follows.

$$G_{\mathrm{En}}(\mathbf{r}) = \frac{|E_{\mathrm{loc}}(\omega_0,\mathbf{r})|^2}{|E_0(\omega_0)|^2} \cdot \frac{|E_{\mathrm{loc}}(\omega_{\mathrm{R}},\mathbf{r})|^2}{|E_0(\omega_{\mathrm{R}})|^2} \tag{S2}$$

, where $E_{\mathrm{loc}}(\omega, \mathbf{r})$ is the local electric-field amplitude at frequency $\omega$ at position $\mathbf{r}$, and $E_0(\omega)$ is the corresponding incident field amplitude at the same frequency. Here, $\omega_0$ denotes the incident frequency and $\omega_{\mathrm{R}}$ denotes the frequency of the Raman-scattered light. Since the Raman shift is typically much smaller than the incident frequency, $\omega_{\mathrm{R}}$ can be well approximated as $\omega_0$. Accordingly, for a plasmonic nanostructure, the field-enhanced Raman emission measured at position $\mathbf{r}$ can be approximated as follows.[3]

$$I_{\mathrm{En}}(\mathbf{r}) \approx \frac{|E_{\mathrm{loc}}(\omega_0,\mathbf{r})|^4}{|E_0(\omega_0)|^4} \times I_{\mathrm{ref}} \propto |E_{\mathrm{loc}}(\omega_0, \mathbf{r})|^4 \tag{S3}$$